\documentclass[reprint,amsmath,amssymb,aps,pre,superscriptaddress]{revtex4-1} 

\usepackage[svgnames]{xcolor}
\usepackage{tikz} 
\usepackage{bm} 
\usepackage{xspace}
\usepackage{subcaption} 
\usepackage{dcolumn}
\usepackage[capitalise,noabbrev]{cleveref} 
\usepackage{siunitx}  

\crefname{equation}{}{} 
\Crefname{equation}{Equation}{Equations}

\newcommand{\mytitle}{Customer mobility and congestion in supermarkets}

\newcommand{\eg}{e.g.\xspace}
\newcommand{\ie}{i.e.\xspace}

\newcommand{\etal}{et al.\xspace }

\begin{document}
\title{\vspace{-10ex}\mytitle} 
\author{Fabian Ying}
\affiliation{Mathematical Institute, University of Oxford, Oxford OX2 6GG, UK}
\author{Alisdair O. G. Wallis}
\affiliation{Tesco PLC, Tesco House, Shire Park, Kestrel Way, Welwyn Garden City, AL7 1GA, UK}
\author{Mariano Beguerisse-D\'{i}az}
\affiliation{Mathematical Institute, University of Oxford, Oxford OX2 6GG, UK}
\author{Mason A. Porter}
\affiliation{Department of Mathematics, University of California, Los Angeles, California 90095, USA}
\author{Sam D. Howison}
\affiliation{Mathematical Institute, University of Oxford, Oxford OX2 6GG, UK}
\date{\today}


\begin{abstract}

The analysis and characterization of human mobility using population-level mobility models is important for numerous applications, ranging from the estimation of commuter flows in cities to modeling trade flows between countries. However, almost all of these applications have focused on large spatial scales, which typically range between intra-city scales to inter-country scales. In this paper, we investigate population-level human mobility models on a much smaller spatial scale by using them to estimate customer mobility flow between supermarket zones. We use anonymized, ordered customer-basket data to infer empirical mobility flow in supermarkets, and we apply variants of the gravity and intervening-opportunities models to fit this mobility flow and estimate the flow on unseen data. We find that a doubly-constrained gravity model and an extended radiation model (which is a type of intervening-opportunities model) can successfully estimate 65--70\% of the flow inside supermarkets. Using a gravity model as a case study, we then investigate how to reduce congestion in supermarkets using mobility models. We model each supermarket zone as a queue, and we use a gravity model to identify store layouts with low congestion, which we measure either by the maximum number of visits to a zone or by the total mean queue size. We then use a simulated-annealing algorithm to find store layouts with lower congestion than a supermarket's original layout. In these optimized store layouts, we find that popular zones are often in the perimeter of a store. Our research gives insight both into how customers move in supermarkets and into how retailers can arrange stores to reduce congestion. It also provides a case study of human mobility on small spatial scales.

\end{abstract}

\maketitle



\section{Introduction}
\label{sec:introduction}

Understanding human mobility is important for city planners, policy makers, transportation researchers, and many others. Motivated by practical applications and the desire to explore fundamental phenomena in human sciences, many researchers have developed and analyzed population-level models, such as gravity \cite{zipf1946p} and intervening-opportunities (IO) models \cite{stouffer1940intervening,schneider1959gravity}, to study human mobility \cite{barbosa2018human}. 

Population-level mobility models characterize the flow of people or other entities between locations using local attributes, such as their populations or the distance between the locations. They have been used for many applications, including modeling commuter flow between locations \cite{simini2012universal}, trade flow between countries \cite{bergstrand1985gravity}, and traffic flow inside a city \cite{piovani2018measuring}. 
    These applications are diverse, but they are all on large spatial scales, ranging from an inter-country scale of thousands of kilometers (\eg, estimating trade flow \cite{anderson1979theoretical,bergstrand1985gravity}) to city and regional scales of tens of kilometers (\eg, estimating commuting patterns \cite{masucci2013gravity}). For even smaller spatial scales (\eg, building level), the prevalent approach in most studies is to use pedestrian models, such as mobility models for individuals (\eg, random walks \cite{gutierrez2015active}) or models of crowd dynamics \cite{helbing1995social}.

    We consider the problem of modeling mobility flow between zones inside supermarkets and investigate how the flow changes when we rearrange store layouts. We therefore examine aggregate flow, which (despite the small spatial scales of these systems) makes population-level mobility models more suitable than random walks~\footnote{However, as we show in \Cref{sec:mobility_models}, we can interpret population-level mobility models as flows that arise from random walks.} or crowd-dynamic models. The small, building-level spatial scale may affect the fundamental features of mobility dynamics (and therefore the performance of the models) in important ways.
    For example, it has been reported that some models (such as radiation models) perform worse on small spatial scales than on large ones \cite{masucci2013gravity}. A possible reason for this observation is that the spatial `force' is much smaller on small scales than on large ones due to the smaller cost of making a trip in the former situation, so other non-spatial `forces' that are not captured by these models may instead be the primary factors that underlie the flow. In a supermarket, for example, the distance between two zones (the spatial `force') may be less relevant than the number of their complementary items (a non-spatial `force') for the flow between the two zones.
    Furthermore, these models are inherently memoryless, as they describe mobility flow from an origin location to a destination location using local attributes of the origin and destination locations, without considering the location from which (or how) a person who leaves the origin location entered it
    in the first place. When one models humans walking inside a building (\eg, in a supermarket or a museum), the direction from which a person comes likely influences where that person goes next, so there is memory in the system. For example, Farley and Ring \cite{farley1966stochastic} observed that customers in supermarkets tend to move from the entrance unidirectionally along the outer perimeter after entering a store.

    In the present paper, we conduct a detailed case study of mobility models in an investigation of congestion in supermarkets, a practical problem that is influenced by the layout of a store. Reducing congestion is important not only for improving the shopping experience of customers, but also for reducing the fulfillment time and cost of online orders. (In many supermarkets, staff members go around a store alongside customers and pick up items that were ordered online.)
    Congestion may delay such orders and thereby incur additional costs to a business and inconvenience customers in a store.
    In our study, we integrate mobility models with a congestion model --- in which each supermarket zone is a queue and in which we make the simplifying assumption that customers traverse shortest paths between purchases --- to estimate congestion in supermarkets. We then use a simple optimization algorithm to find store layouts with low congestion.

    Our article has three main contributions.
    First, we show that several different mobility models can successfully estimate the majority of observed trips in supermarket customer-flow data, demonstrating that these models can work on small (specifically, building-level) spatial scales.
    Second, we show how to combine these models with a congestion model based on queuing networks to estimate congestion in customer flow.
    Third, we demonstrate how to optimize store layouts to reduce congestion.

    Our article proceeds as follows.
    In \Cref{sec:mathematical_set_up}, we describe our mathematical setup. 
    In \Cref{sec:data}, we describe the data set from which we infer the origin--destination (OD) trips in $17$ supermarkets.
    In \Cref{sec:mobility_models}, we describe the mobility models and goodness-of-fit measures that we use in our investigation. We also describe how we estimate the parameters of our models.
    In \Cref{sec:results}, we present our results when applying these models to supermarket store data, using both (in-sample) fitting and (out-of-sample) estimation of customer flows.
    In \Cref{sec:optimization}, we describe an application of a human mobility model to estimate customer congestion and determine store layouts that reduce it.
    Specifically, we discuss our congestion model, our optimization method, and the results of the optimization.
    We conclude and discuss future research directions in \Cref{sec:discussion_and_conclusion}, and we give some additional details about our work in appendices.


\section{Mathematical setup} 
\label{sec:mathematical_set_up}

    In this section, we set up our approach for analyzing mobility flow in supermarkets. We discuss how we discretize space in a supermarket, 
    how we model shopping journeys, and how we characterize flow between zones of a supermarket. We will discuss our data in Section \ref{sec:data} and mobility models in Section \ref{sec:mobility_models}.

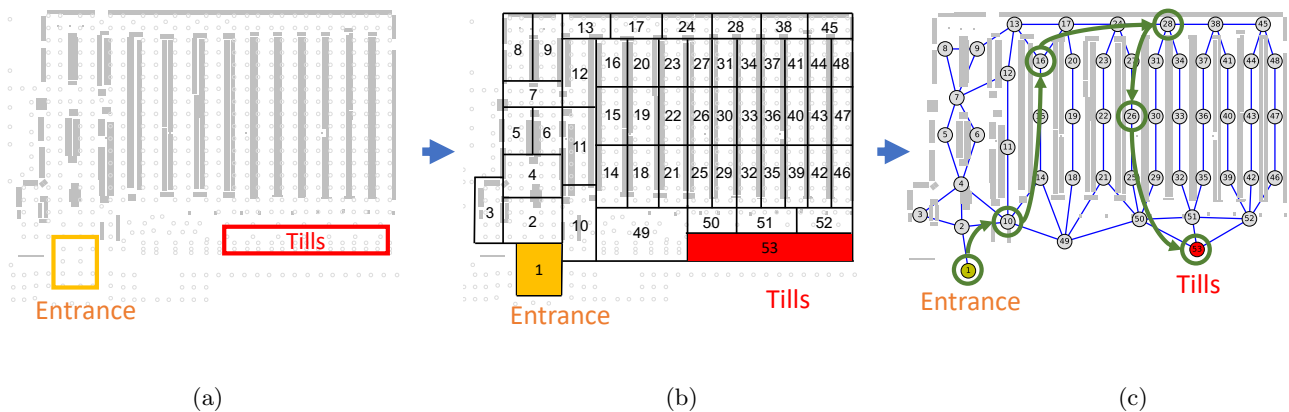
\begin{figure*}
    \begin{tikzpicture}
        \node[anchor=south west,inner sep=0] at (0,0) {\includegraphics[width=\textwidth]{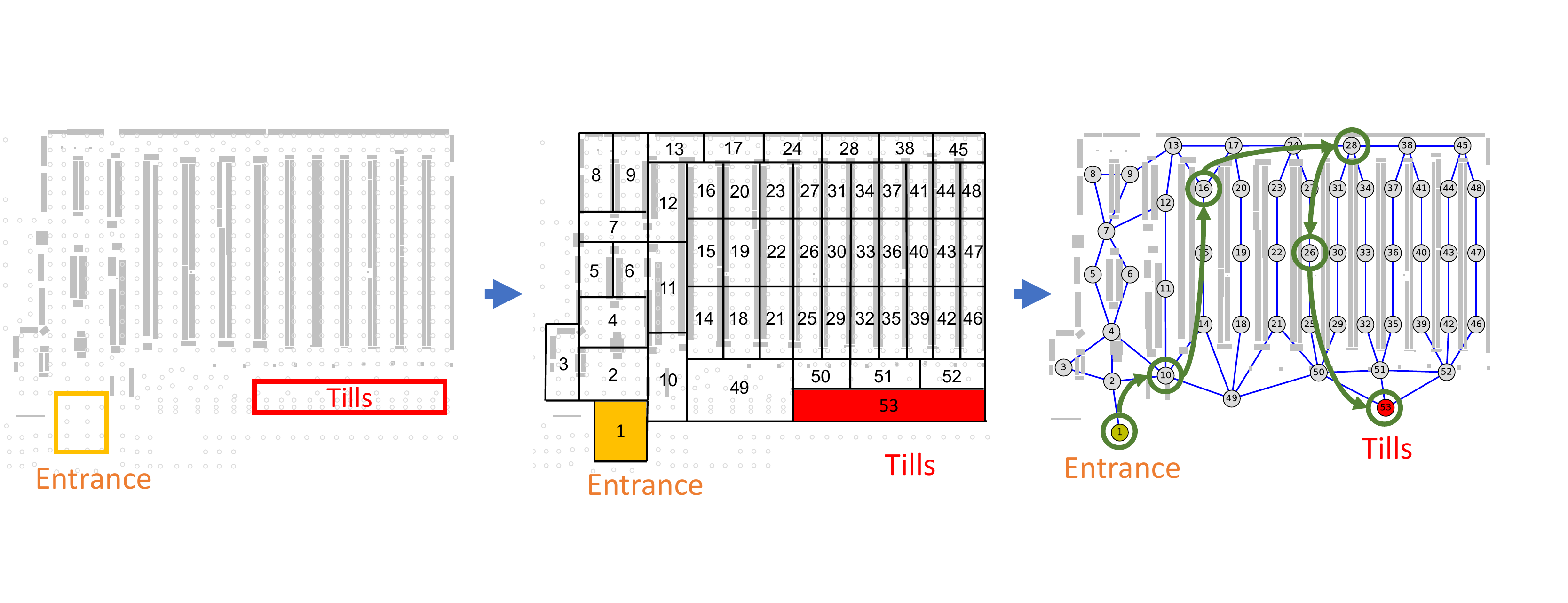}};
        \node[] at (2.7,0.25) {(a)};
        \node[] at (9,0.25) {(b)};
        \node[] at (15,0.25) {(c)};
      \end{tikzpicture}
       \caption{We divide (a) a store into (b) zones and represent it as (c) a network. 
       We depict the shelves in gray. We map the journey, which we highlight in green in panel (c), of a customer who purchases an item (e.g., bread) in zone 10, another item (e.g., milk) in zone 16, a third item (e.g., butter) in zone 28, and a fourth item (e.g., pasta) in zone 26 to a shopping journey (1, 10, 16, 28, 26, 53) and then divide it into 5 origin--destination (OD) trips: (1, 10), (10, 16), (16, 28), (28, 26), and (26, 53). Each green arrow in panel (c) represents an OD trip.}
       \label{fig:store_graph_transformation2_with_trips}
    \end{figure*}

    In our investigation, we employ mobility models that require us to discretize space (\ie, a supermarket), which we divide into a discrete number of disjoint locations, with an associated measure of distance between distinct locations.
    To do this, we manually divide each store into rectangular zones of approximately equal size. (See Appendix~\ref{app:store_zoning} for details.)
    We then represent a store as a network $\mathcal{G}$ with $n$ nodes (representing the zones) and $m$ edges, which connect neighboring zones (see \Cref{fig:store_graph_transformation2_with_trips}). 
    We distinguish an entrance zone (labeled $1$) and a till zone (labeled $n$).
    A store network $\mathcal{G}$, which is embedded in space, is undirected. 
    Although there are distances between supermarket zones, the network $\mathcal{G}$ itself is unweighted.
    For the location of each node, we use the centroid of its corresponding zone.
    For each edge $(i,j)$, we assign an edge length $l_{ij}$, which we take to be the Euclidean distance between its two incident nodes $i$ and $j$. 
    (The edge length approximates the walking distance between two nodes \footnote{A definition of edge length that is closer to the walking distance between two nodes is the length of a shortest path that does not cross any shelves between the centroids of zones $i$ and $j$. 
    We call this distance the ``shelf-respecting shortest-path distance''. 
    However, the difference between the Euclidean distance and shelf-respecting shortest-path distance is small (in most cases, the straight line between the centroids of two zones only clips a corner of a shelf), so we use the former because it is considerably faster to compute.}.)
    We define an $n \times n$ distance matrix $\bm{\Lambda}$ that is associated with $\mathcal{G}$. The entry $d_{ij}$ of $\bm{\Lambda}$ is equal to the \emph{shortest-path distance} between $i$ and $j$; this distance is the minimum length of a path between $i$ and $j$.
    We define the \emph{zone length} of each zone as the length of the longer side of the rectangle that encloses the zone.

    One customer's \emph{shopping journey} is a sequence of $K+2$ zones ($s_0,\dots,s_{K+1}$), where $K$ is the number of items that the customer buys,  $s_0 = 1$ (entrance), $s_{K+1} = n$ (tills), and $s_1,\dots,s_K$ are the zones at which a customer picks up items, which we order by their pick-up times.  
    A customer can purchase multiple items in the same zone, so $s_0,\dots,s_{K+1}$ may not be distinct.
    Each consecutive pair $(s_k, s_{k+1})$ of distinct zones (so $s_k \neq s_{k+1}$) for $0 \leq k \leq K$ constitutes an \emph{origin--destination (OD) trip} (or simply a \emph{trip}). 
    That is, a trip is a segment of a customer's shopping journey that is either between consecutive purchases in different zones, from the entrance to the first purchase, or from the last purchase to the tills.
    We are interested in the number $T_{ij}$ of OD trips 
    in a store from each origin zone $i$ to each destination zone $j$ (over some duration $\tau$).
    We do not consider flow within a zone and thus set $T_{kk} = 0$ for $k = 1,\dots,n$. The $n \times n$ matrix $\bm{T}$, with entries $T_{ij}$, is called an \emph{origin--destination (OD) matrix} \cite{barbosa2018human}; its off-diagonal entries record mobility flow between zones. 
    We denote an empirical OD matrix by $\bm{T}^{\mathrm{data}}$ and an OD matrix from a model by $\bm{T}^{\mathrm{model}}$.
    Throughout our paper, we denote an origin node of a trip by $i$ and a destination node of a trip by $j$; we index other nodes using the symbol $k$.
    We summarize our main notation in \Cref{tab:table_of_notations}.

    \begin{table}[ht]
    \caption{Summary of notation.}
    \label{tab:table_of_notations}
    \begin{ruledtabular}
    \begin{tabular}{l l}
    Variable & Description\\
    \colrule
    $\mathcal{G}$ & Store network\\
    $\bm{\Lambda}$ & Distance matrix (with entries $d_{ij}$) associated with $\mathcal{G}$ \\
    $l_{ij}$ & Length of edge $(i,j)$ in the graph $\mathcal{G}$\\ 
    $d_{ij}$ & Shortest-path length between $i$ and $j$ in the graph $\mathcal{G}$\\
    $l$ & Mean zone length (excluding entrance and tills)\\
    $n$ & Number of nodes in the graph $\mathcal{G}$ \\
     $m$ & Number of edges in the graph $\mathcal{G}$ \\
     $\tau$ & Time period \\
     $\rho$ & Fraction of baskets in the data set\\
            & (i.e., the number of baskets in the data set in \\ & a time period divided by the total number \\ & of baskets 
            during the same time period)\\
    $T_{ij}$ & Number of origin--destination (OD) trips \\
    &  from zone $i$ to zone $j$\\
    $\bm{T}$ & OD matrix with entries $T_{ij}$\\
    $O_k$ & Number of OD trips that start at zone $k$\\
          & (it equals the sum of the entries in the $k^{\text{th}}$ row of $\bm{T}$)\\
    $D_k$ & Number of OD trips that terminate at $k$\\
          & (it equals the sum of the entries \\ & in the $k^{\text{th}}$ column of $\bm{T}$)\\
    $f_{ij}$ & Attraction factor of zone $j$ to zone $i$\\
    $S_{ij}$ & Number of intervening opportunities \\ & 
    of origin--destination pair $(i,j)$ \\ &
              (it equals $\sum_{\{k: d_{ik} < d_{ij}\}} D_k$)\\
    $N$ & Number of trips 
    (it equals $\sum_{i,j} T_{ij}$)\\
    $C$ & Number of shopping journeys \\
        & (it equals $O_1 - D_1$)\\
    $\mu_k$, $\mu$ & Service rate of zone $k$ \\ & (we use $\mu$ when it is independent of $k$)\\
    $v_k$ & Estimated number of visits to zone $k$\\
    $\lambda_k$ & Rate of customer arrival 
    at zone $k$\\
    $Q$ & Total mean queue size\\
    $\lambda_{\max}$ & Maximum arrival rate\\
    \end{tabular}
    \end{ruledtabular}
    \end{table}


\section{Data} 
\label{sec:data}

    We use anonymized, ordered customer-basket data from $17$ large stores of
    a major United Kingdom supermarket chain (Tesco) over a common
     three-month period ($91$ days).
     The data consists of a fraction $\rho \approx 0.07$ of all customer baskets in these stores.
     We summarize the properties of the data in \Cref{tab:stores}.
    
    For each store, we infer the number $T_{ij}^{\mathrm{data}}$ of OD trips from zone $i$ to zone $j$ over the $\tau=91$ days from the data as follows. 
    Each ordered customer basket is a list of item purchases, which we order by pick-up time.
    We use item-location data to map each ordered list of purchases to their associated zones in a supermarket. 
    For example, we map a list of purchases (e.g., bread, milk, butter, and pasta) to its corresponding shopping journey (1, 10, 16, 28, 26, 53), where bread is in zone 10, milk is in zone 16, butter is in zone 28, pasta is in zone 26, and the tills are in zone 53 (see \Cref{fig:store_graph_transformation2_with_trips}(c)). In this example, each item has a unique item location, so we can recover the corresponding shopping journey in a straightforward way. 
    However, about 10\% of the purchased items have unknown item locations and about 8\% of the purchased items have multiple item locations.
    We refer to the latter items as \emph{multi-located items}, and we remove items with unknown item locations from customer baskets.
    For each basket with one or more multi-located items, we consider all combinations of possible purchase locations for those items.
    (For example, there are $2^r$ combinations for a basket with $r$ multi-located items with $2$ locations each.)
    For each combination, we calculate the sum of the shortest-path distances between the locations of consecutive purchases in the basket.
    We then choose a combination of the purchase locations that minimizes this sum.
    We do not possess data to validate that customers tend to buy multi-located items at locations that minimize the sum of shortest-path distance between consecutive purchases, but this assumption likely has only small impact on the results of our analysis (see \Cref{sec:results} for these results), as only about 8\% of the items are multi-located.
    
    We decompose each customer shopping journey into its sequence of OD trips and estimate the total number $T_{ij}^{\mathrm{data}}$ of trips from zone $i$ to zone $j$ by counting all OD trips $(i,j)$ from the data. (For example, the previous example shopping journey (1, 10, 16, 28, 26, 53) consists of 5 trips: (1, 10), (10, 16), (16, 28), (28, 26), and (26, 53).) Assuming that the observed mobility patterns in our data set are representative of the mobility of all customers, we rescale $T_{ij}^{\mathrm{data}}$ by multiplying it by $1/\rho$ (where $\rho$ is the number of baskets in the data set in a time period divided by the total number of baskets during the same time period) to estimate the mobility flow of all customers who visit a store.

    \begin{table}[ht]
    \caption{Summary of our data set, which comes from $17$ Tesco stores. 
    For each quantity, we give the minimum, mean, and maximum values across the $17$ stores.
    }
    \label{tab:stores}
    \begin{ruledtabular}
    \begin{tabular}{p{3.8 cm}lll}

     & Min & Mean & Max\\
    \colrule
    Number of zones ($n$)  & 61 & 123 & 197\\
    Number of edges ($m$) & 128 & 236 & 401\\
    Number of baskets & 2479 & 13672 & 29201 \\
    {Mean zone length, excluding entrance and tills ($l$)} & 6.42\,\si{\metre} & 6.91\,\si{\metre} & 7.65\,\si{\metre}\\ 
    \end{tabular}
    \end{ruledtabular}
    \end{table}


\section{Mobility models} 
\label{sec:mobility_models}

    We examine several mobility models, which we use to estimate $\bm{T}^{\mathrm{data}}$.
    Let $O_k^{\mathrm{data}} = \sum_j T_{kj}^{\mathrm{data}}$ and $D_k^{\mathrm{data}} = \sum_i T_{ik}^{\mathrm{data}}$, respectively, be the empirical numbers of trips that depart from and arrive at each node $k$.
    (Note that $O_k^{\mathrm{data}}$ and $D_k^{\mathrm{data}}$ are also the row and column sums, respectively, of $\bm{T}^{\mathrm{data}}$.)
    We consider a class of models that yield an $n \times n$ OD matrix $\bm{T}^{\mathrm{model}}$ from $O_k^{\mathrm{data}}$, $D_k^{\mathrm{data}}$, the store network, and either one or zero fitting parameters. 
    The goal of the models is for $\bm{T}^{\mathrm{model}}$ to be ``close'' to an empirical OD matrix $\bm{T}^{\mathrm{data}}$. 
    (We discuss diagnostics for comparing $\bm{T}^{\mathrm{model}}$ and $\bm{T}^{\mathrm{data}}$ in \Cref{sub:goodness_of_fit_measures}.)
    The models use $2n$ pieces of information of $\bm{T}^{\mathrm{data}}$ to estimate the $(n-1)\times(n-1)$ off-diagonal entries of $\bm{T}^{\mathrm{data}}$.
    
    In our problem, $O_k^{\mathrm{data}} = D_k^{\mathrm{data}}$ for all nodes $k$ except for $k=1$ (entrance) and $k=n$ (tills), as every customer who finishes a trip in zone $k$ (except for $k = 1$ and $k = n$) continues their journey with a trip that starts from $k$.
    {Note that $O_1^{\mathrm{data}} \neq D_1^{\mathrm{data}}$, because each shopping journey starts at node $1$, so the first trip (which starts at $1$) of each shopping journey does not have a preceding trip that ends at $1$.
    Similarly, $O_n^{\mathrm{data}} \neq D_n^{\mathrm{data}}$, because each shopping journey ends at node $n$, so the last trip (which ends at $n$) does not have a subsequent trip that starts at $n$.}
    The quantity $O_k^{\mathrm{data}}$ also corresponds to the number of \emph{shopping visits} at $k$ (except for $k = 1$ and $k = n$). It is thus equal to the total number of times that customers visit $k$ to purchase one or more items.
    The number $C$ of journeys in the data satisfies both $C = O_1^{\mathrm{data}} - D_1^{\mathrm{data}}$ and $C = D_n^{\mathrm{data}} - O_n^{\mathrm{data}}$, as every journey starts at node $1$ and ends at node $n$.
    Note that $D_1^{\mathrm{data}}$ and $O_n^{\mathrm{data}}$ need not be equal to $0$, as the entrance and till nodes can contain items.
    Therefore, one is able to determine $\{O_k^{\mathrm{data}}\}_{k=1}^n$ and $\{D_k^{\mathrm{data}}\}_{k=1}^n$ from $\{O_k^{\mathrm{data}}\}_{k=1}^n$ and $D_1^{\mathrm{data}}$.
    In practice, we estimate these values from sales data (see Appendix~\ref{app:shopping_visits}).
    Therefore, we assume that we know $\{O_k^{\mathrm{data}}\}_{k=1}^n$ and $\{D_k^{\mathrm{data}}\}_{k=1}^n$.

    We use \textit{doubly-constrained} (also called \textit{production--attraction-constrained}) versions \cite{wilson1971family} of mobility models.
    In these models, the mobility flow $\bm{T}^{\mathrm{model}}$ satisfies
    \begin{align}
        \label{eq:O_k_conservation}
         O_k^{\mathrm{data}} &= O_k^{\mathrm{model}}\,, \\
         D_k^{\mathrm{data}} &= D_k^{\mathrm{model}} \,,
         \label{eq:D_k_conservation}
    \end{align}
    where $O_k^{\mathrm{model}} = \sum_j T_{kj}^{\mathrm{model}}$ and $D_k^{\mathrm{model}} = \sum_i T_{ik}^{\mathrm{model}}$.
    In other words, $\bm{T}^{\mathrm{model}}$ has the same row sums and column sums as $\bm{T}^{\mathrm{data}}$. This, in turn, implies for each node $k$ that both the number of trips that arrive at $k$ and the number of trips that depart from $k$ are equal to their empirical values.
    Therefore, for notational simplicity, we drop the superscripts on $O_k$ and $D_k$ for the remainder of our paper.
    Because $O_k = D_k$ for $k=2,\dots,n-1$, the number of people at each node (except for the entrance and till nodes) is also conserved in the models.

    For each origin node $i$, there is a vector of `attraction values' $f_{ij}$ for each possible destination $j$.
    We calculate these values from a model-specific function $f_{\mathrm{model}}$ that takes $O_i$, $\{D_k\}_{k=1}^n$, and information (such as the distance between two nodes) from a store network as inputs.
    The function $f_{\mathrm{model}}$ is the same for each origin node $i$.
    (Allowing this function to be heterogeneous for different nodes would allow us to incorporate different types of supermarket zones into our models.)
    The mobility flow in a doubly-constrained model is
    \begin{equation}
        T_{ij}^{\text{model}} = (A_i O_i) \times (B_j D_j) \times f_{ij}\,,
        \label{eq:base_model}
    \end{equation}
    where $A_i, B_j \geq 0$ are ``balancing factors'' to ensure that \Cref{eq:O_k_conservation,eq:D_k_conservation} are satisfied.
    Given $O_k,$ $D_k$, and $f_{ij}$, we determine $A_i$ and $B_j$ using an iterative proportional-fitting procedure \cite{deming1940least}. See Appendix~\ref{app:iterative_proportional_fitting} for more details.

    We can interpret $T_{ij}^{\text{model}}$ as the mean aggregate flow that arises from a continuous-time random-walk model at stationarity.
    In our models, customers arrive at node $1$ (i.e., the entrance) at a rate of $\lambda = C / \tau$.
    In contrast to a standard random walk on a network \cite{masuda2017random}, customers do not choose a random neighbor 
    at each step.
    Instead, each customer at node $i$ chooses a random destination $j$, which need not be adjacent to $i$, with probability $P_{ij} = A_i B_j D_j f_{ij}$, and it then takes a trip from $i$ to the chosen destination $j$. That is, the customer traverses some path from $i$ to the chosen destination $j$.
    For simplicity, we assume that customers take a shortest path from node $i$ to node $j$, where we choose this path uniformly at random from all shortest paths between these two nodes. 
    (Other routing models are possible; one possibility is a standard random walk that starts at node $i$ and reaches an absorbing state at node $j$.)
    We remove customers who finish a trip at node $n$ at rate $\lambda$; this ensures that the mean number of customers in the system 
    is constant.
    The quantity $T_{ij}^{\text{model}}$ thus gives the mean number of trips from zone $i$ to zone $j$ over a period $\tau$.
    Our model assumes that there is no memory in customer mobility; the next destination of each customer depends only on its present location. 

    We present each model in a scale-invariant form, such that the parameters are dimensionless and the attraction values $f_{ij}$ are invariant under the scalings $O_k \mapsto a O_k$ and $D_k \mapsto a D_k$ for $a > 0$ and for all $k$.
    Because $O_k$ and $D_k$ scale with the number $C$ of journeys in the data set (and therefore with $\tau$), scale invariance ensures that the model parameters and the transition probabilities $P_{ij}$ are independent of $C$ and $\tau$~\footnote{For sufficiently large $\tau$, it would be interesting to generalize our analysis to incorporate seasonal shopping variations, such as differences in behavior during holidays.}.
    
    In \Cref{tab:mobility_models}, we summarize the different choices of $f_{ij}$ and the number of parameters for each of the four models that we employ. We discuss these models in the following subsections.


    \subsection{Gravity models}

        Gravity models of mobility \cite{carey1867principles,zipf1946p,wilson1970entropy,erlander1990gravity}, which are named after
        Newton's law of gravity, have been used to model a variety of systems, including human migration \cite{erlander1990gravity,jung2008gravity,lee2014matchmaker}, cargo-ship movement \cite{kaluza2010complex}, inter-city telecommunication flow \cite{krings2009urban}, spatial accessibility of health services \cite{luo2003measures}, and trade flow \cite{anderson1979theoretical,bergstrand1985gravity}. In a gravity model, the mobility flow between two locations depends only on the distance between the locations and on the `mass' (i.e., `population') of the two locations. In our models, each node $k$ (i.e., zone) has two types of populations: there is an origin population $O_k$ (the number of trips that depart from zone $k$) and a destination population $D_k$ (the number of trips that arrive at zone $k$). We use the origin population when we calculate the mobility flow from $k$ (\ie, the outflow from $k$) and the destination population when we calculate the mobility flow to $k$ (\ie, the inflow to $k$). Because $O_k = D_k$ for $k \in \{2, \dots, n-1\}$ in our problem, the two values for population are the same, except for the entrance ($k=1$) and till ($k=n$) nodes.

        We use a doubly-constrained gravity model, with
        \begin{equation}
              f_{ij} = f_{\mathrm{g}}(O_i, D_j, d_{ij}) = O_iD_j\left(d_{ij}/l\right)^{-\gamma}\,,
          \label{eq:gravity}
        \end{equation}
        where $l > 0$ is a spatial normalization factor (which we choose to be the mean zone length, excluding the entrance and till zones) and $\gamma \geq 0$ is a dimensionless fitting parameter. 
        We exclude the entrance and till zones in calculating mean zone length, because they are typically much larger than the other zones in a store. 
        See Appendix~\ref{app:store_zoning} for more details.

        The expression $\left(d_{ij}/l\right)^{-\gamma}$ is an example of a `deterrence function', for which an exponential function is also a common choice \cite{barbosa2018human}. In contrast to other studies on small spatial scales \cite{lenormand2016systematic,liang2013unraveling,simini2012universal}, we find that a power-law deterrence function gives a (slightly) better fit to our data than an exponential deterrence function (see Appendix~\ref{app:exp_deterrence}). 


    \subsection{Intervening-opportunities models} 
    \label{sub:IO_model}
 
        In intervening-opportunities (IO) models, which were first proposed by Stouffer in 1940 \cite{stouffer1940intervening}, 
        each location (\ie, node) has opportunities, which (depending on their number and/or quality) give the location some amount of `popularity'.
        The key concept of IO models is the notion of \textit{intervening opportunities}.
        The intervening opportunities $S_{ij}$ 
        of an OD pair $(i,j)$
        consist of 
        all opportunities in nodes $k$ 
        that satisfy $d_{ik} < d_{ij}$. 
        (Note that $S_{ij} \neq S_{ji}$ in general.)
        In IO models, the mobility flow between two locations (\ie, zones in our application) depends on the number of intervening opportunities (rather than on the distance) between the two locations and on the `populations' of the two locations.
        A larger number of intervening opportunities of an OD pair $(i,j)$ entails a smaller mobility flow from zone $i$ to zone $j$, because customers are more likely to find what they are looking for (or to be diverted) before they reach $j$. 
        Intervening-opportunities models and their variants have been used in many applications, including to model intra-city mobility \cite{ruiter1967toward}, interstate migration \cite{anderson1955intermetropolitan,akwawua2000intervening,akwawua2001development}, rioting behavior \cite{davies2013mathematical}, and the creation of social ties in a city \cite{sim2015great}.

        We measure the number of opportunities at each node~$k$ by $D_k$, the number of trips that arrive at $k$.
        The number $S_{ij}$ of intervening opportunities of an OD pair $(i,j)$
         is then
        \begin{equation}
            S_{ij} = \sum_{\substack{k \neq i \\ d_{ik} < d_{ij}}} D_k\,.
        \end{equation}         
        Therefore, the `opportunities' in our problem amount to
        opportunities for customers to stop and purchase something.
        Note that $S_{ij} \mapsto a S_{ij}$ when we scale $D_k \mapsto a D_k$ for all $k$, so the number of opportunities scales linearly with $C$.

        In Stouffer's original IO formulation (StIO), the number of people who move a given distance is proportional to the number of opportunities at that distance and is inversely proportional to the number of intervening opportunities.
        The attraction values $f_{ij}$ are 
        \begin{equation} \label{here}
            f_{ij} = f_{\mathrm{StIO}}(D_j, S_{ij}) = \frac{D_j}{S_{ij} + cN}\,,
        \end{equation}
        for some $c > 0$ (to avoid dividing by $0$) and where $N = \sum_{i,j} T_{ij}$ is the total number of trips.
         In \Cref{here}, we use $cN$ instead of $c$ to ensure scale invariance.
        Additionally, $f_{ij}$ is only approximately inversely proportional to $S_{ij}$ (because of the $cN$ term).
        
        In our investigation, we use Schneider's reformulation of the IO model \cite{schneider1959gravity}, because it is more popular than Stouffer's IO formulation \cite{barbosa2018human} and is underpinned by a mechanistic model. 
        In this reformulation, the attraction values $f_{ij}$ are
        \begin{equation}
            f_{ij} = f_{\mathrm{IO}}(D_j, S_{ij}) =  e^{-\frac{L}{N} S_{ij}} - e^{-\frac{L}{N} (S_{ij} + D_j)} > 0 \,,
            \label{eq:io_model_base}
        \end{equation}
        where $L \in (0, N]$ is a dimensionless fitting parameter.
        The quantity $f_{ij}$ equals the number of customers at node $i$ who take a trip to node $j$ divided by the number of customers who leave $i$ under the following mechanistic model.
        Each customer at $i$ considers opportunities in nondecreasing order of distance from $i$, and they accept each opportunity with probability $L/N$.
        They take a trip to the node $j$ that has the first opportunity that they accept.
        One can show that the number of customers who take a trip from $i$ to $j$ divided by the total number who leave $i$ is equal to the right-hand side of \Cref{eq:io_model_base} (see Appendix~\ref{app:derivation_schneider_IO_model}).
        We use a doubly-constrained version of the IO model, so $f_{ij}$ gives only the attraction value of zone $j$ to a customer in zone $i$. 
        In a doubly-constrained model, the quantity $T^{\text{model}}_{ij} / O_i = A_i B_j D_j f_{ij}$ equals the actual number of customers who make a trip from $i$ to $j$ divided by the total number of trips that originate at $i$.


    \subsection{Radiation model}

        The original radiation model \cite{simini2012universal}, which was proposed as an alternative to gravity models, is a parameter-free variant of the IO model with attraction values 
        \begin{equation} \label{eq:radiation_model}
            \begin{aligned}
        f_{ij} &= f_{\mathrm{rad}}(O_i, D_j, S_{ij}) \\
            &= \frac{O_i D_j}{(O_i + S_{ij})(O_i + D_j + S_{ij})}\,.
        \end{aligned}
        \end{equation}

        The radiation model and its variants have been used for studies of commuter flows \cite{simini2012universal,simini2013human}, human migration \cite{lee2014matchmaker}, mobile-phone calls \cite{simini2012universal}, and other applications.
        An advantage of this version of the radiation model is that it has no parameters. However, it does not appear to do a good job of capturing human mobility on small spatial scales \cite{lenormand2012universal,masucci2013gravity,liang2013unraveling}.


    \subsection{Extended radiation model}

        Yang \etal \cite{yang2014limits} proposed an extension of the radiation model that includes an exponent $\alpha$.
        In this model, the attraction values $f_{ij}$ are
        \begin{widetext}
        \begin{equation}
             f_{ij} 
            = f_{\mathrm{ext}}(O_i, D_j, S_{ij}) 
            = \frac{\left[(O_i + S_{ij} + D_j)^{\alpha} - (O_i + S_{ij})^{\alpha}\right]((O_i)^{\alpha} + N^\alpha)}
                     {((O_i + S_{ij})^{\alpha} + N^\alpha) \left[(O_i + S_{ij} + D_j)^{\alpha} + N^\alpha\right]} \,.
        \end{equation}
        \end{widetext}

        Yang \etal claimed that this extended radiation model fits empirical OD matrices 
        better than the original radiation model for intra-city commuting flow and observed that their calibrated values of $\alpha$ decreased as they considered systems with smaller spatial scales. 
        When $\alpha=1$, one recovers a variation of the original radiation model from \Cref{eq:radiation_model}
        (specifically, with each
         occurrence of $O_i$ replaced by $O_i +  N$).

        { 
        \renewcommand{\arraystretch}{1.5}
        \begin{table*}[ht]
        \caption{Summary of the four mobility models that we employ. The OD matrix of each model is given by \Cref{eq:base_model}, with different functional forms for the attraction values $f_{ij}$. }
        \label{tab:mobility_models}
        \begin{ruledtabular}
        \begin{tabular}{l l l l l}
        Model  & $f_{ij}$ & Parameter & Parameter range & References\\
        \colrule
        Gravity   & $D_j\left(d_{ij}/l\right)^{-\gamma}$ 
        & $\gamma$ & $[0, \infty)$ & \cite{carey1867principles,zipf1946p,wilson1970entropy}
        \vspace{0.5em}\\
        Intervening opportunities (IO)        & $\exp\left(-\frac{L}{N} S_{ij}\right) - \exp\left(-\frac{L}{N} (S_{ij} + D_j)\right)$ 
        & $L$ & $[0, \infty)$ & \cite{schneider1959gravity,ruiter1967toward}
        \vspace{0.5em}\\
        
        Radiation & $\frac{O_i D_j}{(O_i + S_{ij})(O_i + D_j + S_{ij})}$ 
        & - & - & \cite{simini2012universal,masucci2013gravity,simini2013human}
        \vspace{0.5em}\\

        Extended radiation  & $\frac{\left[(O_i + S_{ij} + D_j)^{\alpha} - (O_i + S_{ij})^{\alpha}\right]((O_i)^{\alpha} + N^\alpha)}
                     {((O_i + S_{ij})^{\alpha} + N^\alpha) \left[(O_i + S_{ij} + D_j)^{\alpha} + N^\alpha\right]}$ 
            & $\alpha$ & $[0, \infty)$ & \cite{yang2014limits} \\
        \end{tabular}
        \end{ruledtabular}
        \end{table*}
        } 


    \subsection{Goodness-of-fit measures}       
    \label{sub:goodness_of_fit_measures}
    
    
        \subsubsection{Common part of commuters (CPC)}

            The \emph{common part of commuters} (CPC) score is the proportion of trips that the OD matrices $\bm{T}^{\text{data}}$ and $\bm{T}^{\text{model}}$ have in common:
            \begin{equation}
                \mathrm{CPC}(\bm{T}^{\text{data}}, \bm{T}^{\text{model}}) 
                = \frac{\sum_i \sum_j \min\{T^{\text{data}}_{ij}, T^{\text{model}}_{ij}\}}
                {\sum_i \sum_j \frac{1}{2} (T^{\text{data}}_{ij} + T^{\text{model}}_{ij})}\,.
                \label{eq:CPC}
            \end{equation}
            It was introduced in \cite{gargiulo2012commuting,lenormand2012universal} and has been used in studies of human mobility \cite{yang2014limits,lenormand2016systematic,mcneill2017estimating}.
            The CPC score is based on the S{\o}rensen index \cite{sorensen1948method}, and it varies from $0$ (when there is no agreement between the model and data) to $1$ (when $\bm{T}^{\text{data}}$ and $\bm{T}^{\text{model}}$ are identical). 
            Because our models are doubly-constrained, $\sum_{i,j} T^{\text{data}}_{ij} = \sum_{i,j} T^{\text{model}}_{ij}$. Therefore, we interpret the CPC score as the fraction of customers whose trip is
            assigned correctly by a model.

            There are various other goodness-of-fit measures, such as normalized root-mean-squared error in $\bm{T}$, information gain, common part of edges, cosine distance, and Pearson product-moment correlation.
            However, in past studies, these measures often gave similar results as CPC when comparing the performance of mobility models \cite{lenormand2016systematic,mcneill2017estimating}, so we primarily use CPC, which has an intuitive interpretation in our modeling context.


        \subsubsection{Error in estimated number of zone visits}

            In addition to CPC, we also consider an application-specific goodness-of-fit measure $\mathrm{NRMSE}_v$, which measures the normalized root-mean-square error (NRMSE) in the number of visits to each node.
            When we examine congestion in supermarkets in \Cref{sec:optimization}, we use measures of congestion that depend on the number of visits to each node, so a mobility model should have low values of $\mathrm{NRMSE}_v$ for it to be viable for our application to congestion. 
            Given an OD matrix $\bm{T}$, we estimate the number of visits by assuming that each customer, for an OD trip $(i,j)$, takes a shortest path, which we choose uniformly at random among all shortest paths from $i$ to $j$.
            Each customer who takes a trip from $i$ to $j$ visits each node along the chosen shortest path.
            The estimated number $v_k$ of visits to each node $k$ is the weighted sum of the number $T_{ij}$ of trips with OD pairs $(i, j)$ for all $i$ and $j$, where the weight $\omega_{ikj}$ is the fraction  
            of shortest paths from $i$ to $j$ that traverse $k$.
            (We use the convention that the starting and terminal nodes, $i$ and $j$, are traversed as part of
            a shortest path from $i$ to $j$.)
            That is,
            \begin{equation}
                v_k = v_k(\bm{T}) = \sum_{i,j} \omega_{ikj} T_{ij}\,.
                \label{eq:wbc}
            \end{equation}

            The number $v_k$ of visits is closely related to geodesic node betweenness centrality \cite{newman2018networks}, which we recover when $T_{ij} = 1$ for all $(i, j)$. We can compute all $v_k$ values in $\mathcal{O}(nm)$ time using a straightforward adaptation of a fast algorithm for computing geodesic betweenness centrality \cite{brandes2001faster}. 

            To measure the model error in the estimated number of visits, we calculate the
            NRMSE in $v_k(\bm{T})$ with the formula
            \begin{equation}
                \mathrm{NRMSE}_v = \left(\frac{\sum_{k=1}^n (v_k(\bm{T}^{\mathrm{data}}) - v_k(\bm{T}^{\mathrm{model}}))^2}{n v_{\max}(\bm{T}^{\mathrm{data}})^2}\right)^{\frac{1}{2}}\,,
            \end{equation}
            where $v_{\max}(\bm{T}) = \max_k \left[v_k(\bm{T})\right]$ is the number of visits to the most-visited node.

{Our shortest-path assumption is a modelling choice and seems likely to be somewhat unrealistic for describing the precise trajectories between purchases.
            For example, in one study \cite{hui2009research}, it was reported that customer trajectories between purchases are, on average, about four times as long as a shortest path.
            However, to the best of our knowledge, there does not exist a model that accurately describes how customers move between purchases.
            We use the shortest-path assumption as a null model, and we note that it is possible to replace this assumption with a more intricate model that estimates the number of visits to each node. 
            Additionally, we use the shortest-path assumption only to estimate the number of visits to each node; we do not use this assumption when we estimate OD matrices.}


    \subsection{Parameter calibration}
        \label{sub:parameter_calibration}
  
        Following the approach in \cite{lenormand2012universal}, we calibrate the model parameters $\gamma$, $L$, and $\alpha$ of the gravity, IO, and extended radiation models (respectively) for each data set by maximizing the CPC score. We call a parameter value `optimal' when it
        maximizes the CPC score for a given model and data set.


\section{Results} \label{sec:results}


    \subsection{Fit to data} 
    \label{sub:fit_to_data}

\begin{figure*}
            \begin{subfigure}[b]{0.5\textwidth}
                  \includegraphics[trim={1cm 0.5cm 1cm 1cm},clip,width=\textwidth]{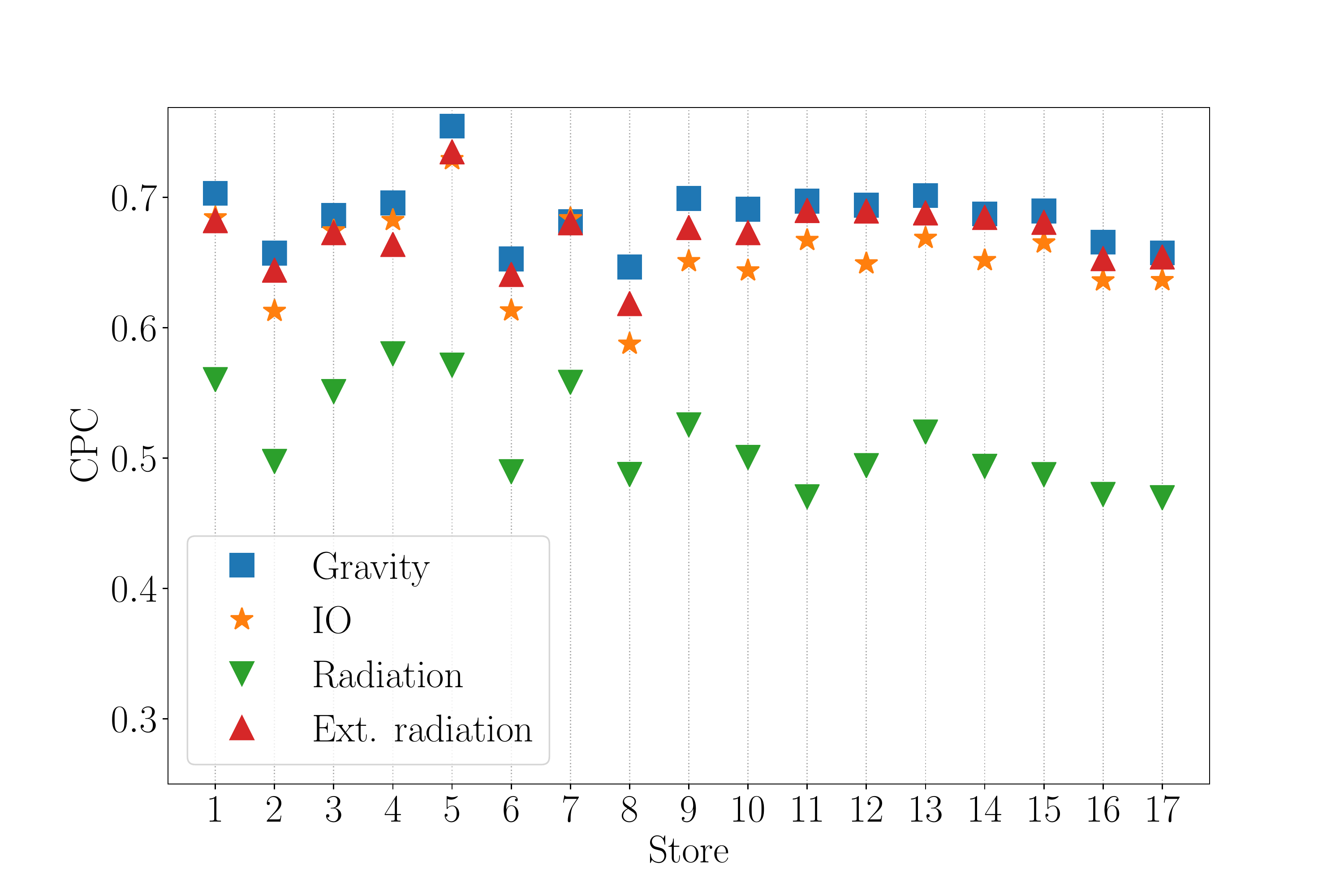}
                  \caption{CPC scores}
                  \label{fig:CPC_results_pl_anonymized_paper.pdf}
            \end{subfigure}%
            \begin{subfigure}[b]{0.5\textwidth}
                  \includegraphics[trim={1cm 0.5cm 1cm 1cm},clip,width=\textwidth]{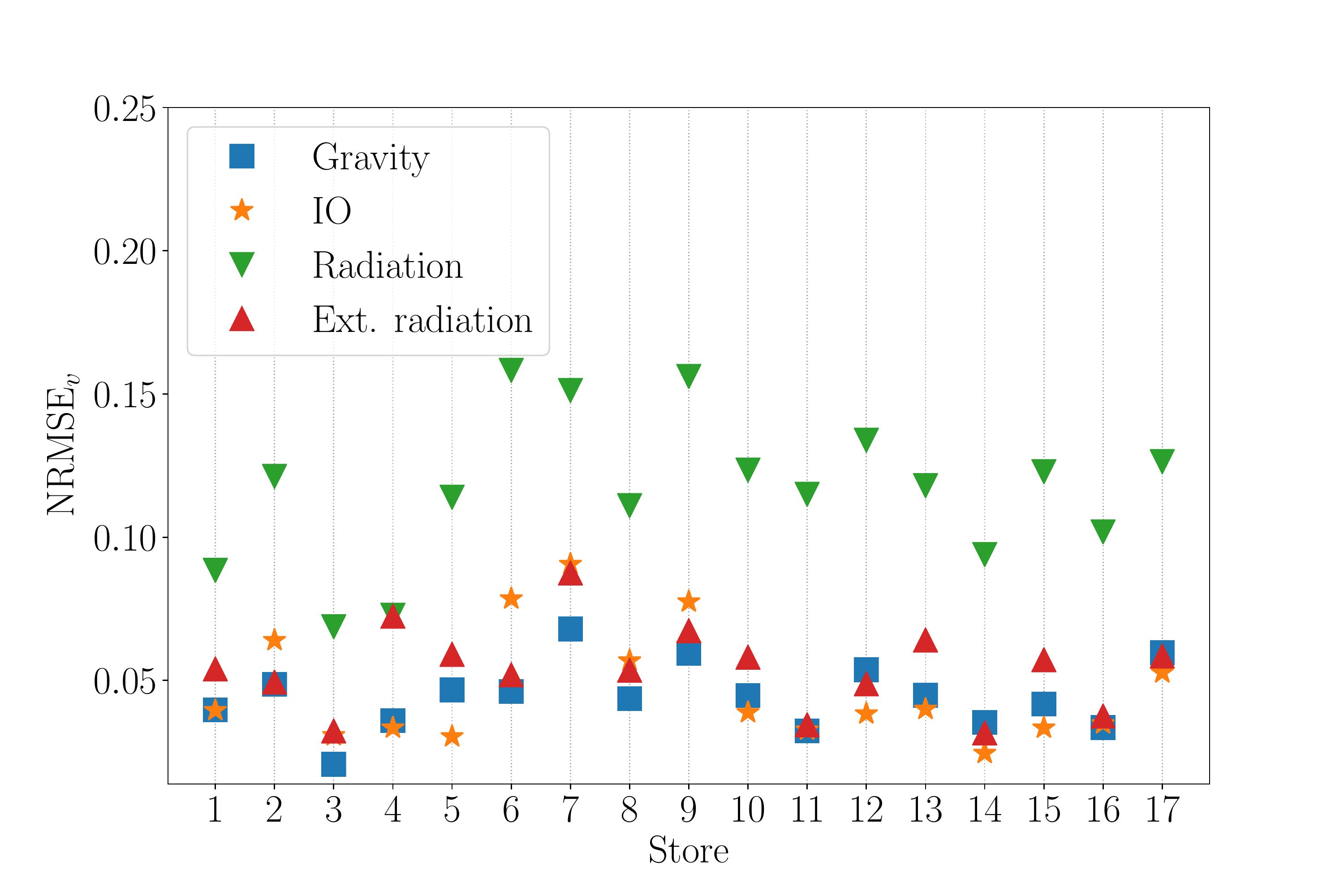}
                  \caption{$\mathrm{NRMSE}_v$}
                  \label{fig:NRMSE_results_pl_anonymized_paper.pdf}
            \end{subfigure}%
            \caption{CPC scores and $\mathrm{NRMSE}_v$ values when fitting the gravity, IO, radiation, and extended radiation models to mobility-flow data from $17$ supermarkets.}
            \label{fig:results_CPC_NRMSE}
        \end{figure*}

        We test the four models in \Cref{sec:mobility_models} on each of the $17$ stores.
        In our computations, the gravity model consistently achieves the best CPC score across the stores, with a mean of about $0.686$ (see \Cref{fig:CPC_results_pl_anonymized_paper.pdf} and \Cref{tab:cpc_rmse}).
        This value is comparable with reported CPC scores in previous studies on mobility systems with larger spatial scales \cite{lenormand2016systematic,mcneill2017estimating,yang2014limits}.
        The performance of the extended radiation model, with a mean CPC score of $0.672$, is almost as successful on average.
        The IO model consistently yields lower CPC scores than the gravity and extended radiation models.
        The gravity model also has the best (\ie, lowest) mean value of $\mathrm{NRMSE}_v$ 
        across the $17$ stores (see \Cref{tab:cpc_rmse}), closely followed by the IO model and then the extended radiation model.
        In terms of $\mathrm{NRMSE}_v$, the relative performance of these three models is store-dependent (see \Cref{fig:NRMSE_results_pl_anonymized_paper.pdf}). In some stores, the gravity model has the lowest value of $\mathrm{NRMSE}_v$; in other stores, either the IO model or the extended radiation model achieves the lowest value.
        For each store, the radiation model performs the worst among the four models, yielding both the lowest CPC scores and the highest values of $\mathrm{NRMSE}_v$ across all 17 stores.
        The poor performance of the radiation model is consistent with other studies of mobility systems on small spatial scales \cite{lenormand2012universal,masucci2013gravity,liang2013unraveling}.

        \begin{table}[ht]
        \caption{Mean CPC scores and $\mathrm{NRMSE}_v$ values from fitting the gravity, IO, radiation, and extended radiation models to mobility-flow data from $17$ supermarkets. We list the models in decreasing order of their mean CPC score.
        We highlight the best value in each column in bold.}
        \label{tab:cpc_rmse}
        \begin{ruledtabular}
        \begin{tabular}{l c c c c}
        Model & Mean CPC & Mean $\mathrm{NRMSE}_v$  \\
        \colrule
        Gravity   & \textbf{0.686} & \textbf{0.045} \\  
        Ext. radiation & 0.672 & 0.054\\
        IO        & 0.655 & 0.047 \\
        Radiation & 0.513 & 0.116 \\
        \end{tabular}
        \end{ruledtabular}
        \end{table}

        To further investigate the performance of the models, we examine the results of a single store (which we call `Store A') in more detail. 
        Specifically, we examine the 
        estimated OD matrix $\bm{T}^{\mathrm{model}}$, the estimated number $v_k(\bm{T}^{\mathrm{model}})$ of visits, and the distance distribution of the OD trips for each of the mobility models.
        The results for this store are qualitatively similar to the results for the other
        stores.

\begin{figure*}
            \begin{subfigure}[b]{0.4\textwidth}
                  \includegraphics[trim={0cm 0 0cm 0},clip,width=\textwidth]{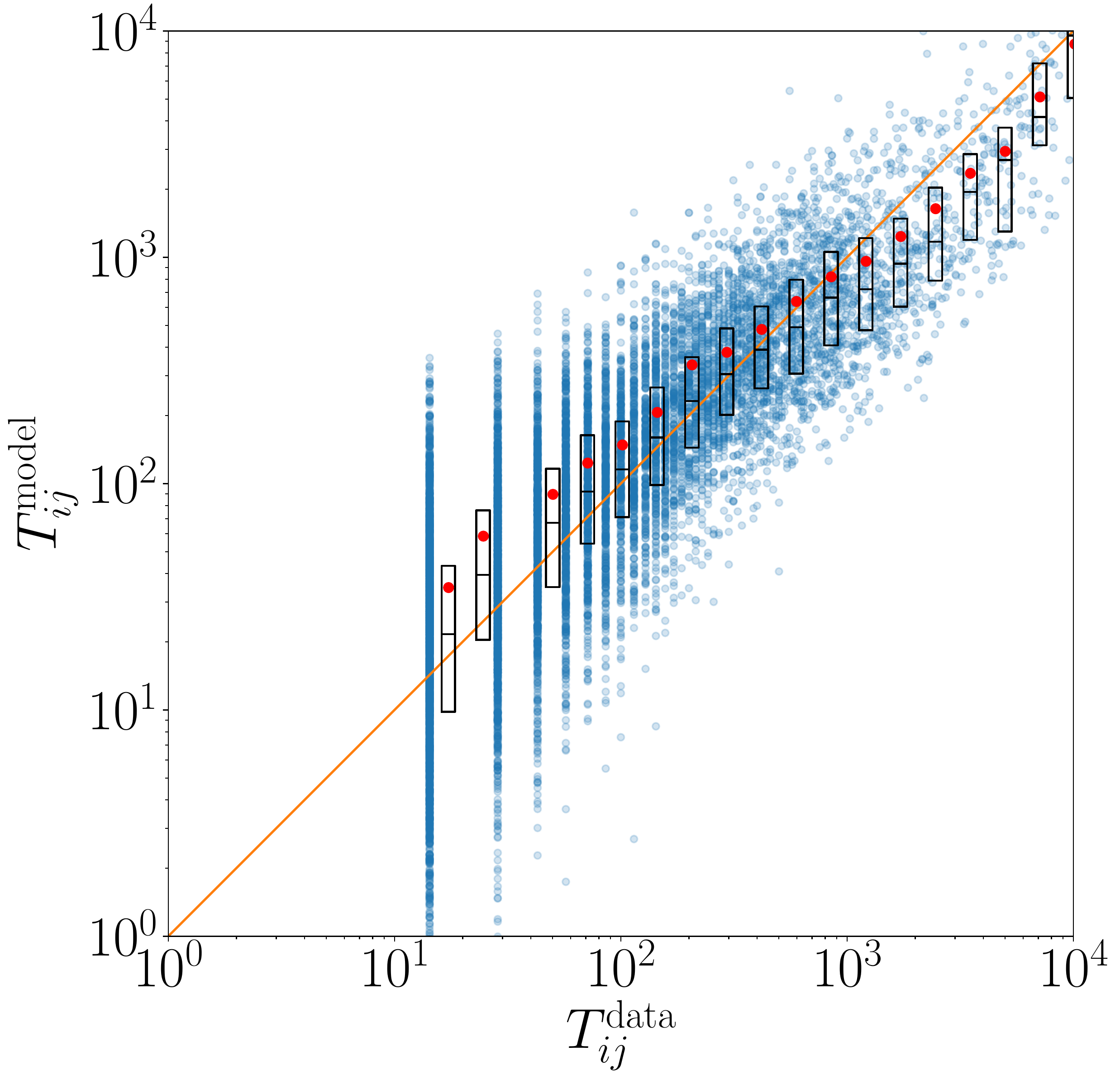}
                  \caption{Gravity model}
                  \label{fig:flux_gravity.pdf}
            \end{subfigure}%
           \hspace{.1 in}
            \begin{subfigure}[b]{0.4\textwidth}
                  \includegraphics[trim={0cm 0 0cm 0},clip,width=\textwidth]{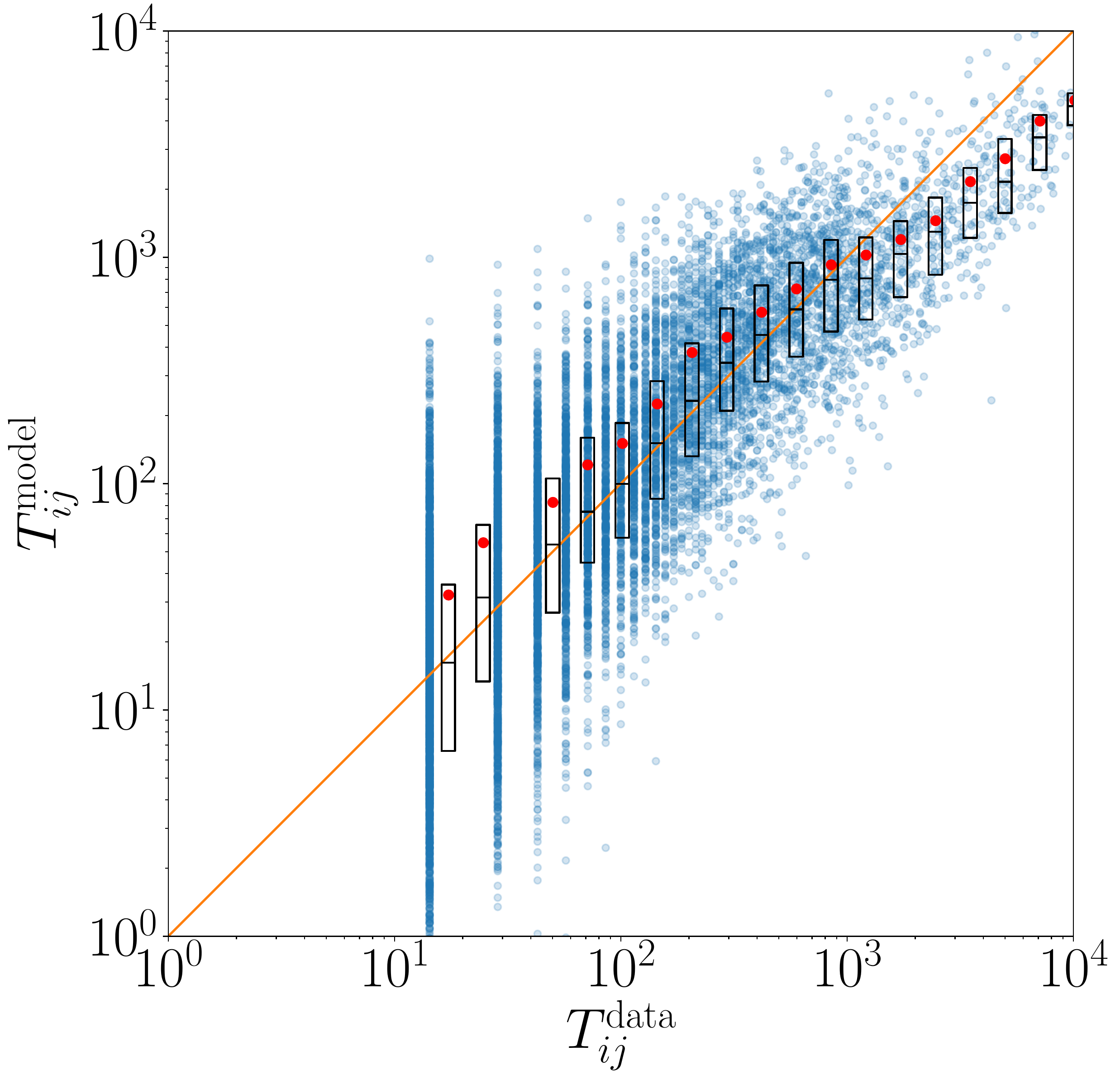}
                  \caption{IO model}
                  \label{fig:flux_IO.pdf}
            \end{subfigure}%
            \par\bigskip %
            \begin{subfigure}[b]{0.4\textwidth}
                  \includegraphics[trim={0cm 0 0cm 0},clip,width=\textwidth]{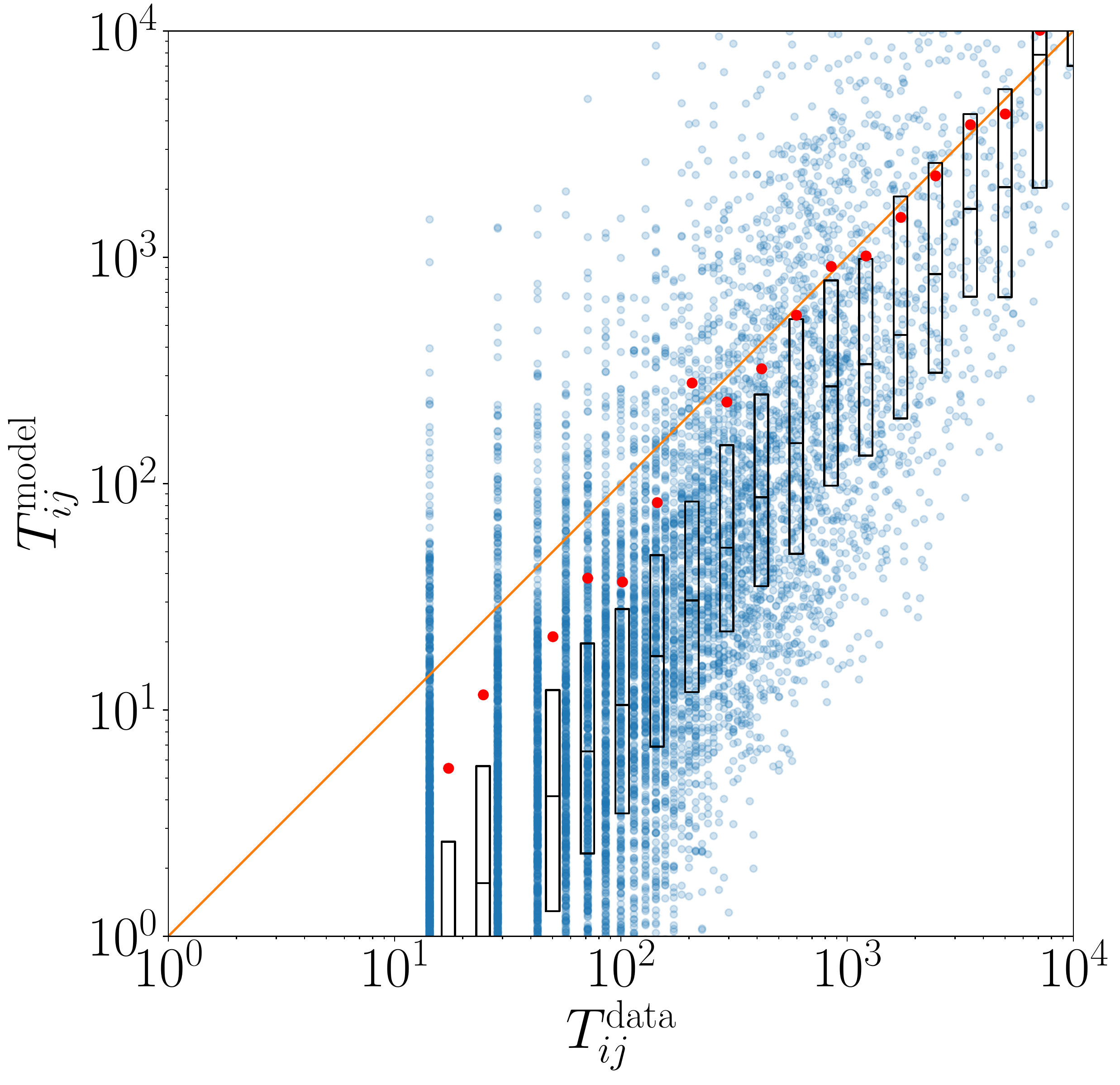}
                  \caption{Radiation model}
                  \label{fig:flux_radiation.pdf}
            \end{subfigure}%
       \hspace{.1 in}
            \begin{subfigure}[b]{0.4\textwidth}
                  \includegraphics[trim={0cm 0 0cm 0},clip,width=\textwidth]{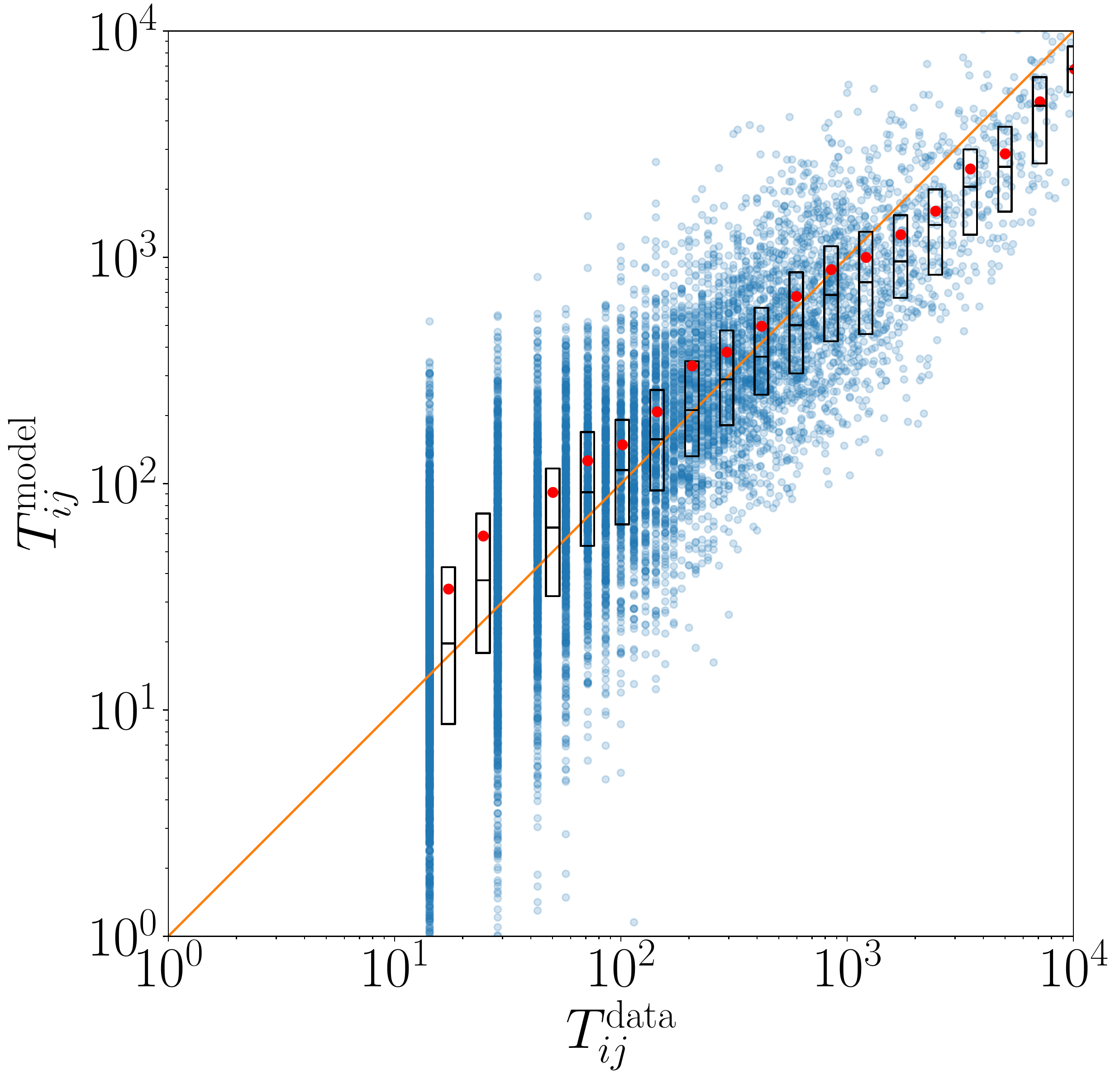}
                  \caption{Extended radiation model}
                  \label{fig:flux_ext-radiation.pdf}
            \end{subfigure}
            \caption{Comparison (blue dots) between the number of trips in
            the data ($T_{ij}^{\mathrm{data}}$) and the model estimation
             ($T_{ij}^{\mathrm{model}}$).
            We also plot the mean number of trips that are estimated
             by the model (red dots) for each logarithmic bin of the data.
            The orange line is the identity line.
            Each box (in black) extends from the lower to the upper quartile values of the binned model estimate;
            we draw a horizontal line at the median.}
            \label{fig:flux_comparison}
        \end{figure*}

        In \Cref{fig:flux_comparison}, we compare the empirical number $T_{ij}^{\mathrm{data}}$ of trips with the estimated number $T_{ij}^{\mathrm{model}}$ of trips from each mobility model for each OD pair $(i,j)$ of nodes in Store A.
        To evaluate the quality of $T_{ij}^{\mathrm{model}}$, we create logarithmic bins from $1$ to $\max_{i,j}(T_{ij}^{\mathrm{data}})$.
        For each bin, we consider all OD pairs
        whose empirical number of trips lies within the bin. We calculate the mean, median, and interquartile range for the estimated number $T_{ij}^{\mathrm{model}}$ of trips for these OD pairs 
        $(i,j)$. 
        See the black box plots in \Cref{fig:flux_comparison}.
    
        On average, the estimated numbers of trips from all four models are close to their empirical numbers, except for OD pairs with a large number of trips. 
        For these OD pairs, the gravity, IO, and the extended radiation models underestimate the number of trips.
       We tested whether this bias results from the presence of
        a large number of very short trip between neighboring shelves on either side of a zone boundary. However, this does not appear to be the case.
        We also examined a more general type of gravity model to explore whether it can improve the fit and/or remove this bias, but it did not.
        The underestimation by the models of the number of trips for OD pairs with many trips 
        remains an interesting and unexplained feature of the empirical data.
          The radiation model is effective at estimating the mean number of trips for most of the bins, but its overall performance is poor because of the large variance in its estimates for each bin (see \Cref{fig:flux_radiation.pdf}).
 
\begin{figure*}
            \begin{subfigure}[b]{0.4\textwidth}
                  \includegraphics[width=\textwidth]{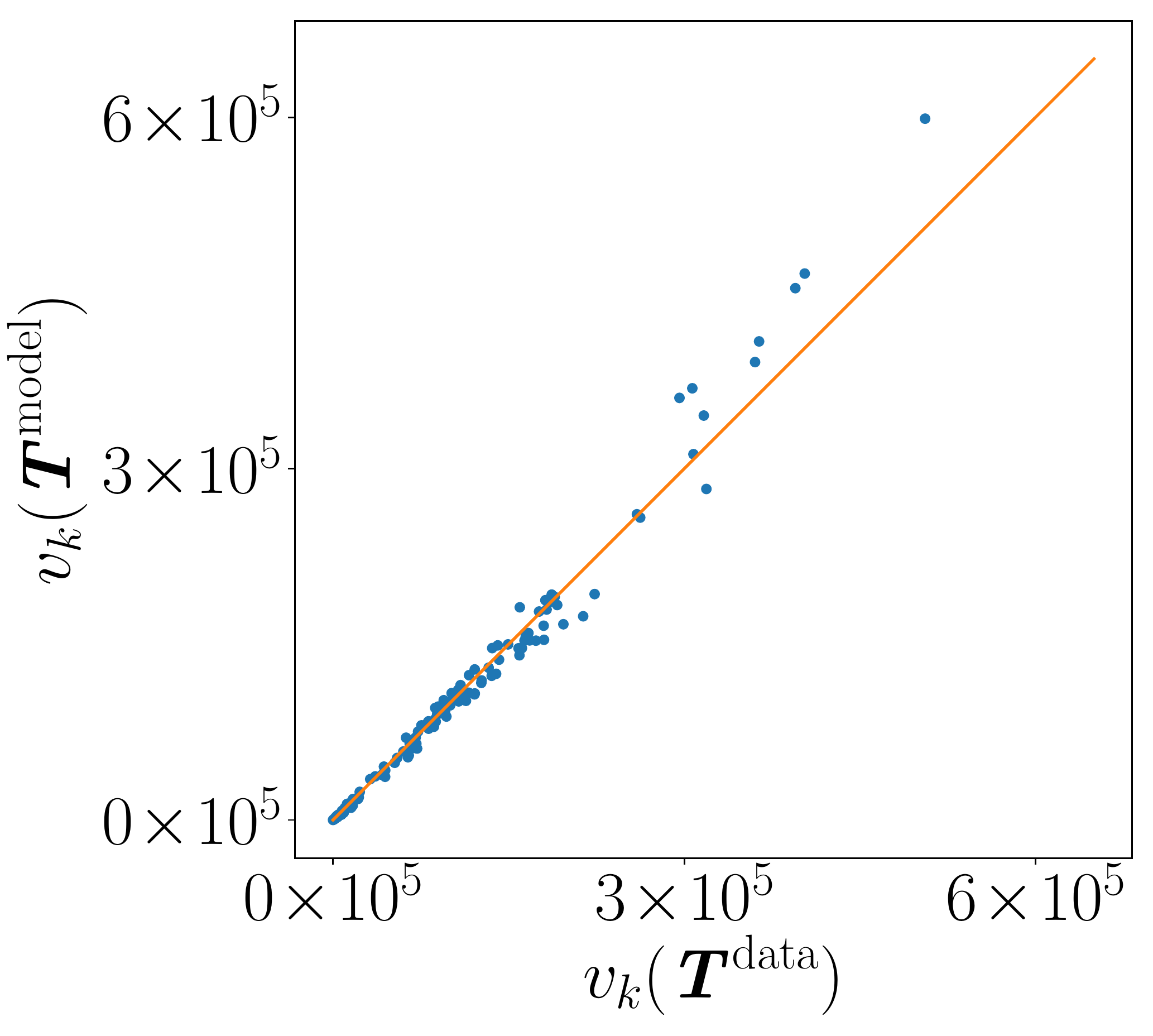}
                  \caption{Gravity model\\ ($\mathrm{NRMSE}_v$: 0.032)\\ \vspace{0.15cm}}
                  \label{fig:num_visits_zone_paper_gravity-pl.pdf}
            \end{subfigure}%
            \hspace{.1 in}
            \begin{subfigure}[b]{0.4\textwidth}
                  \includegraphics[width=\textwidth]{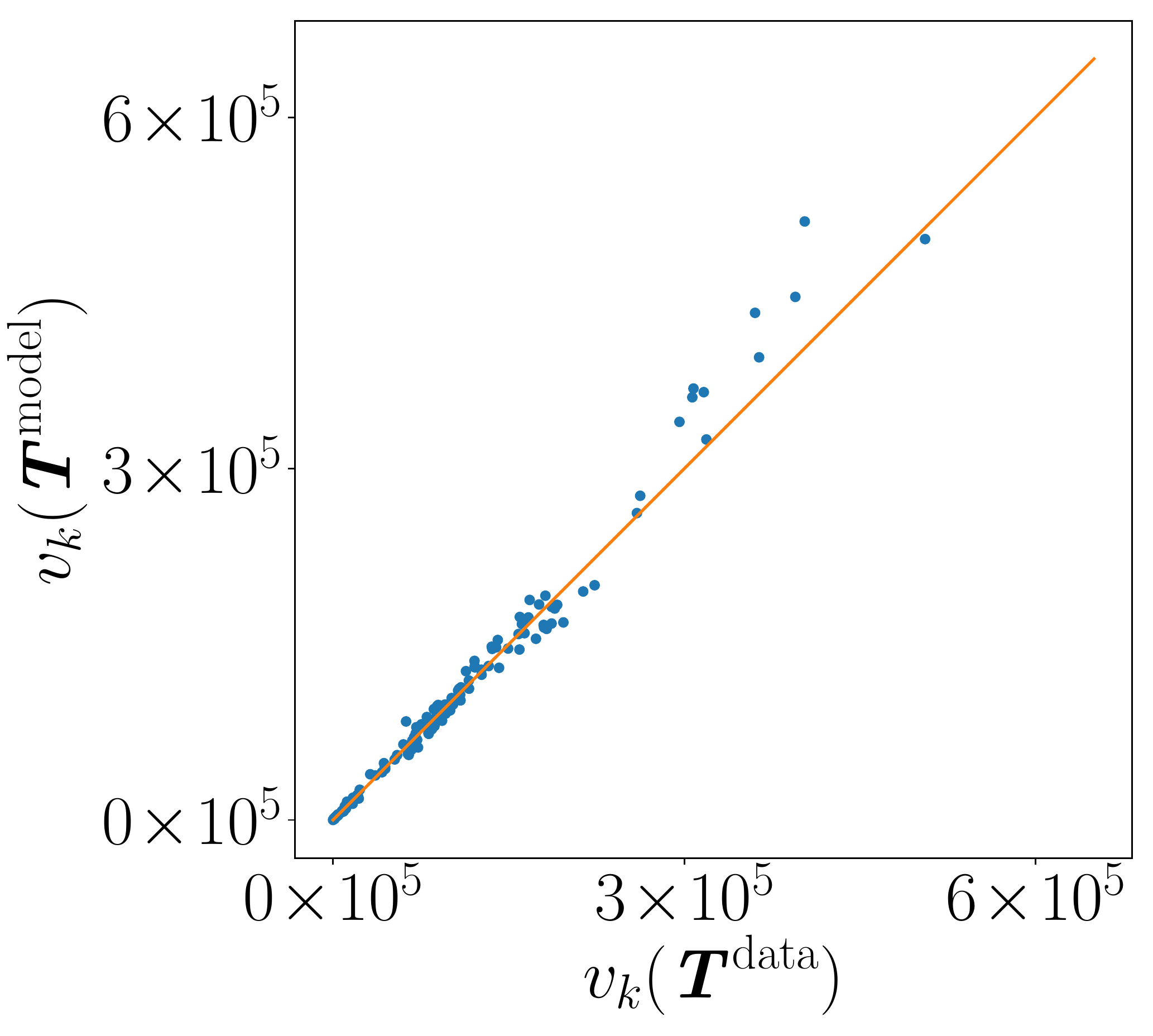}
                  \caption{IO model\\ ($\mathrm{NRMSE}_v$: 0.033)\\ \vspace{0.15cm}}
                  \label{fig:num_visits_zone_paper_IO.pdf}
            \end{subfigure}%
            \par\bigskip %
            \begin{subfigure}[b]{0.4\textwidth}
                  \includegraphics[width=\textwidth]{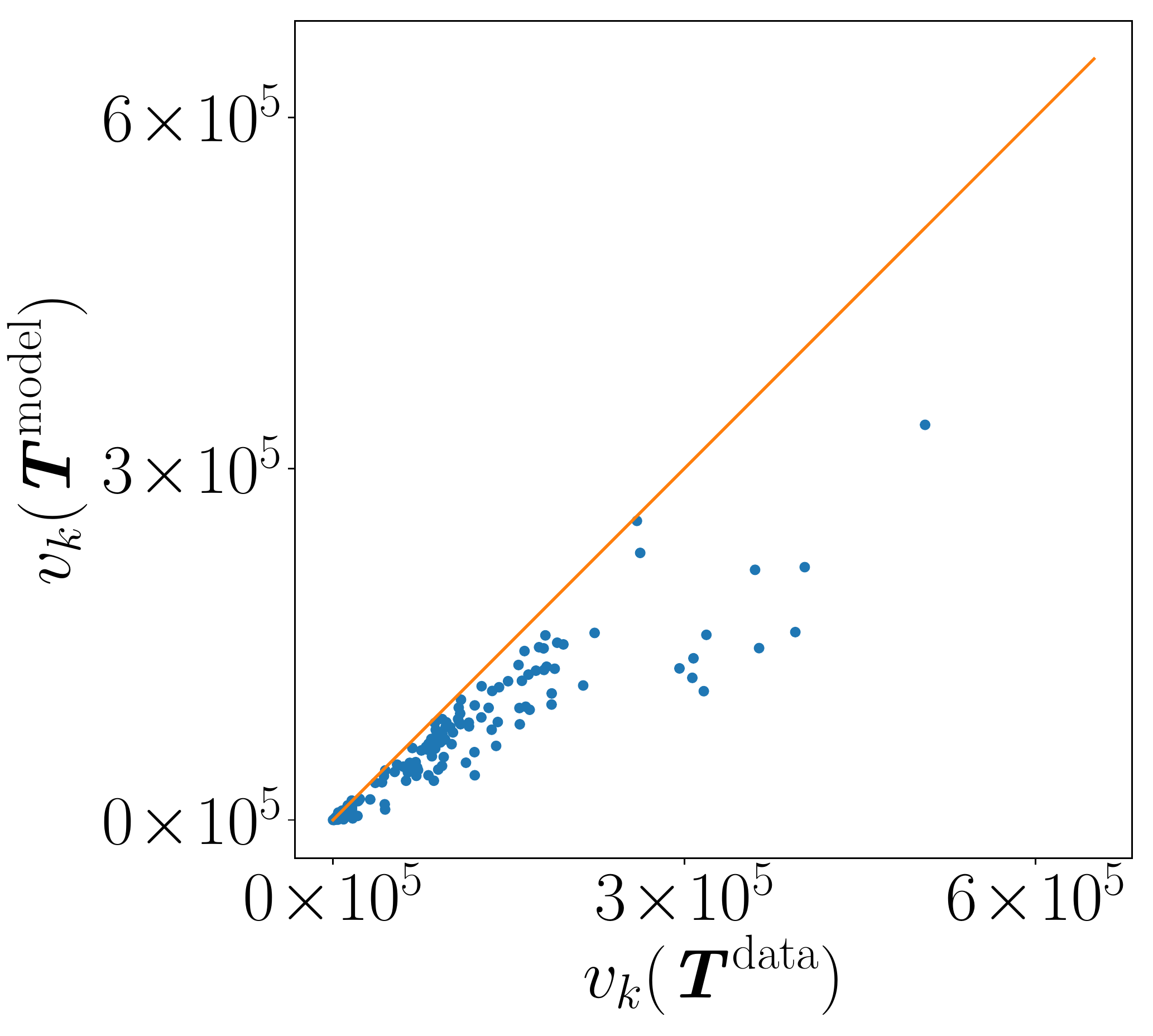}
                  \caption{Radiation model\\ ($\mathrm{NRMSE}_v$: 0.115)\\ \vspace{0.15cm}}
                  \label{fig:num_visits_zone_paper_radiation.pdf}
            \end{subfigure}%
            \hspace{.1 in}
            \begin{subfigure}[b]{0.4\textwidth}
                  \includegraphics[width=\textwidth]{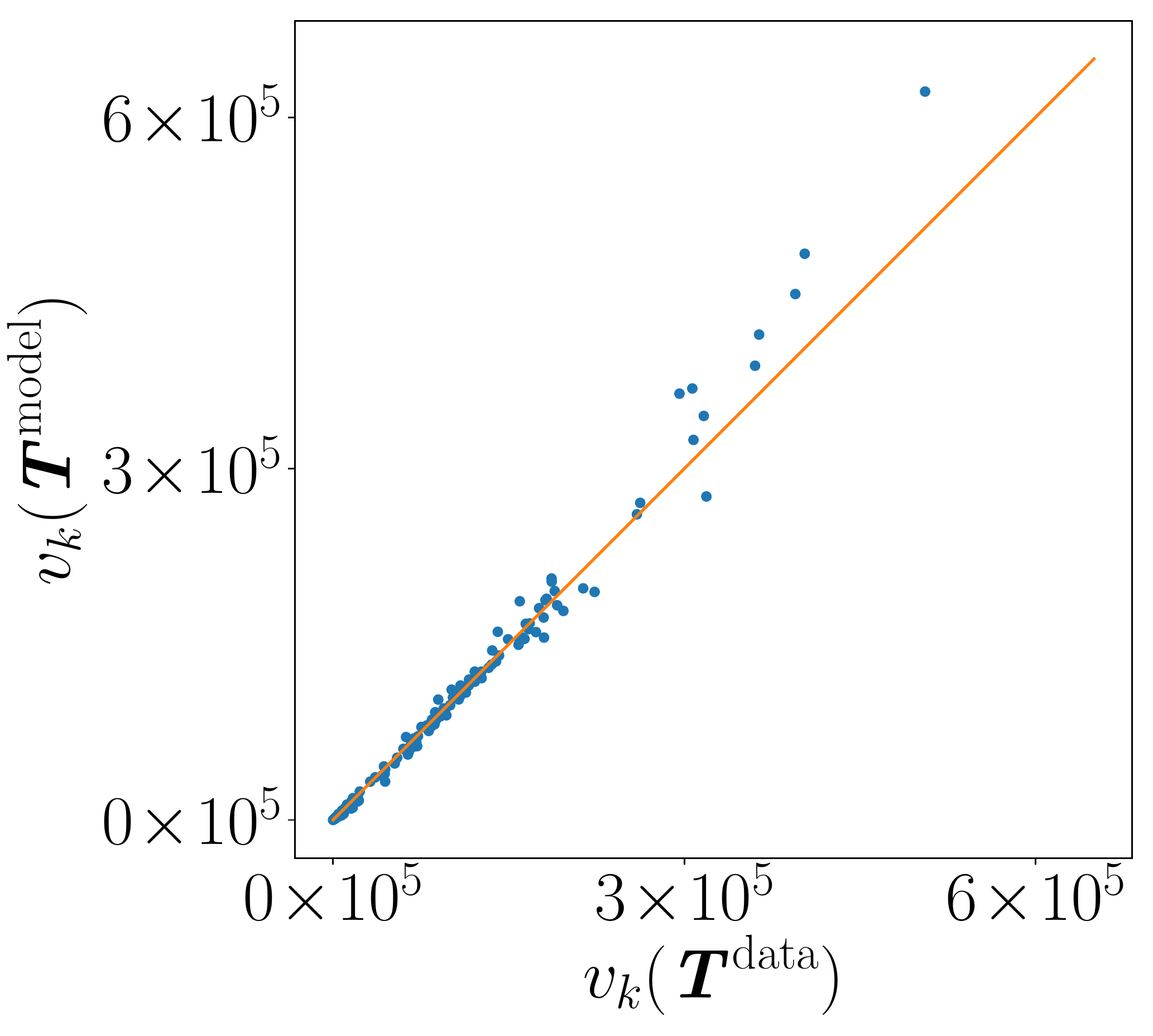}
                  \caption{Extended radiation model\\ ($\mathrm{NRMSE}_v$: 0.034)\\ \vspace{0.15cm}}
                  \label{fig:num_visits_zone_paper_ext-radiation.pdf}
            \end{subfigure}
            \caption{Comparison of the estimated number $v_k$ of visits for each node $k$ 
            between the data and the mobility models. The orange line is the identity line. 
            The gravity, IO, and extended radiation models give good fits to the number of visits to each node.
            }
            \label{fig:num_visits_zone_paper}
        \end{figure*}

         In \Cref{fig:num_visits_zone_paper}, we compare the estimated number $v_k(\bm{T}^{\mathrm{model}})$ of visits that we compute from the OD matrix $\bm{T}^{\mathrm{model}}$ of the models with the number $v_k(\bm{T}^{\mathrm{data}})$ of visits that we estimate using the empirical OD matrix $\bm{T}^{\mathrm{data}}$ for Store A.
        We find that the gravity, IO, and extended radiation models are effective at estimating the number of visits for most nodes, except for some of the ones
         with a large number of visits.
        For these nodes, the three models overestimate the number of visits.
        The radiation model underestimates the number of visits for most nodes (see \Cref{fig:num_visits_zone_paper_radiation.pdf}).

\begin{figure}
              \includegraphics[width=\columnwidth]{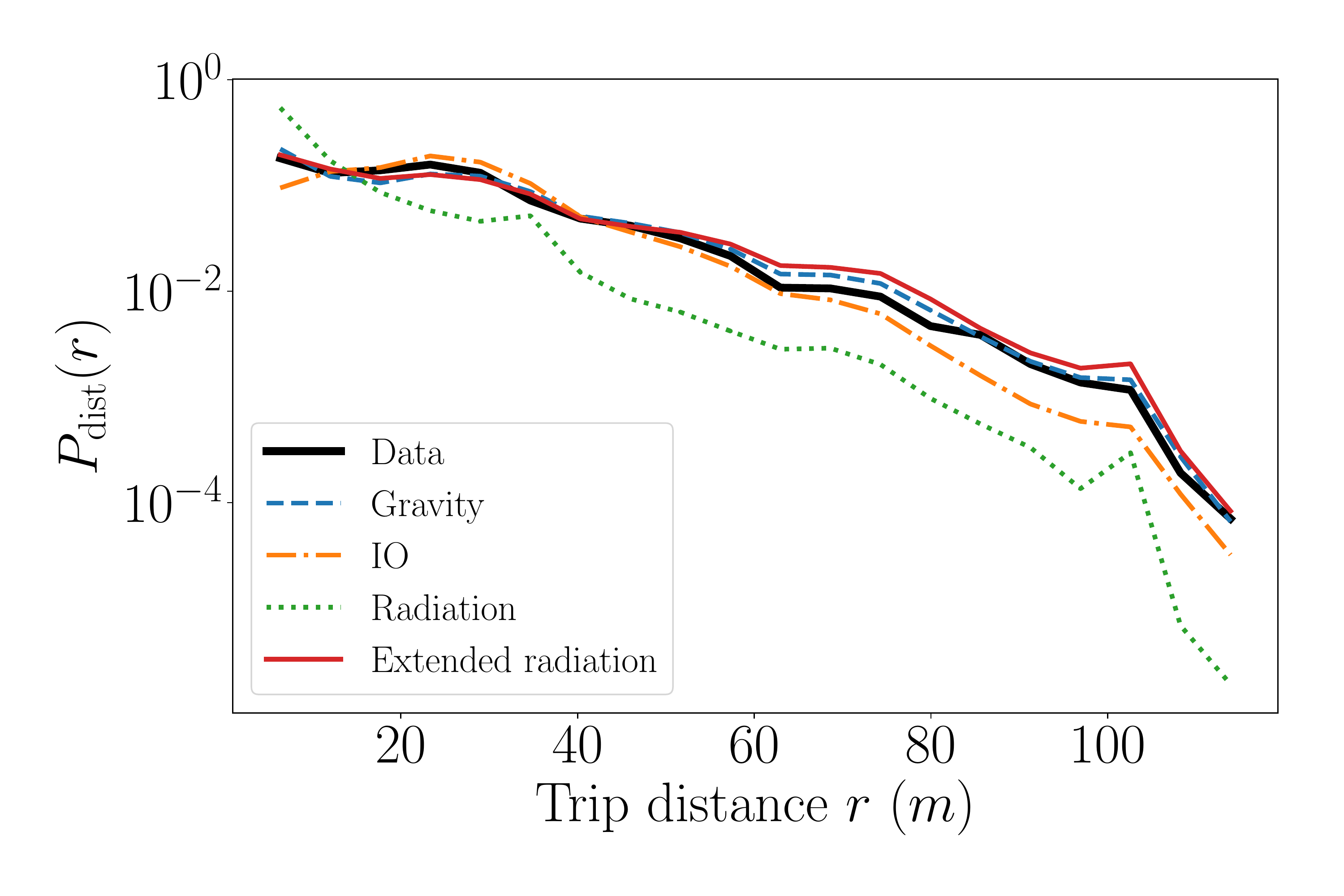}
            \caption{Probability density function (PDF) of the trip-distance distribution.
            }
            \label{fig:distance_distribution}
        \end{figure}

        In \Cref{fig:distance_distribution}, we compare the distribution of trip distances in our models with the empirical distribution. 
        The gravity, IO, and extended radiation models have trip-distance distributions that qualitatively resemble the empirical distribution.
        Among the four models, the trip-distance distribution from the gravity model is closest to the empirical distribution.
        The IO model underestimates the number of long-distance trips (specifically, those above $60\,\si{\metre}$), and the extended radiation model overestimates the number of these long-distance trips.
        The trip-distance distribution of the radiation model is qualitatively different from the empirical distribution (see \Cref{fig:distance_distribution}).

        In summary, among the models that we examine, the gravity model best fits the empirical mobility-flow data.
        On average, it successfully explains about 69\% of the OD trips in the data sets.
        It also is effective at estimating the number of visits to each node, with NRMSE values of about $0.045$ (see \Cref{tab:cpc_rmse}).
        The extended radiation and IO models are close behind; on average, they successfully explain the data of about 65--67\% of the OD trips.
        For the most part, these three models also yield trip-distance distributions that look similar to the empirical distribution.
        The radiation model does not fit the data well, so we exclude it from further consideration.

    \subsection{Sensitivity analysis: Parameter dependence of models}
        \label{sub:sensitivity_analysis}
    
\begin{figure*}[hbt]
            \begin{subfigure}[b]{0.33\textwidth}
                  \includegraphics[trim={1.2cm 1.5cm 1.2cm 1.5cm}, clip, width=\textwidth]{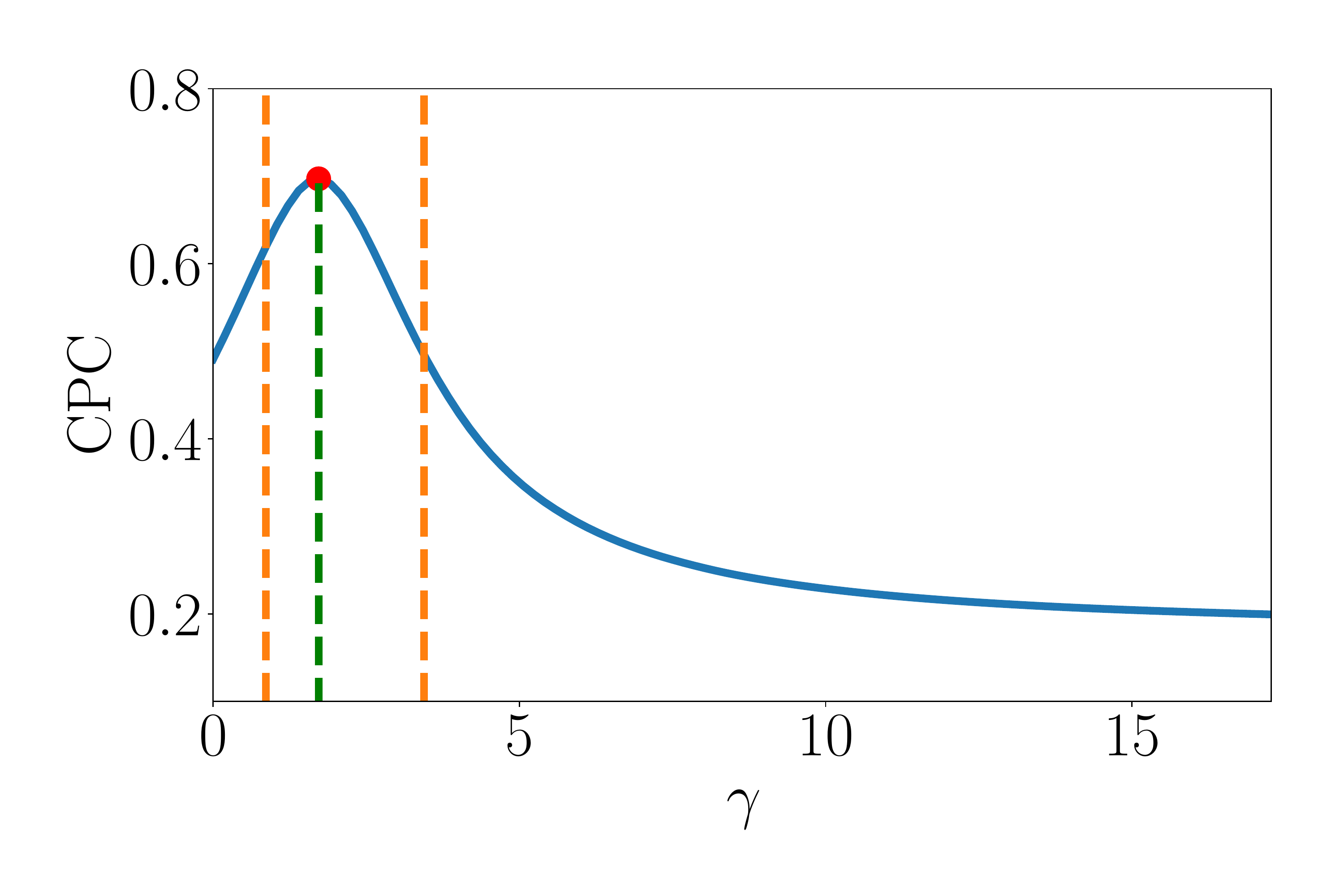}
                  \caption{Gravity model}
                  \label{fig:2050_CPC_vs_parameter_gravity-pl.pdf}
            \end{subfigure}%
            \begin{subfigure}[b]{0.33\textwidth}
                  \includegraphics[trim={1.2cm 1.5cm 1.2cm 1.5cm}, clip, width=\textwidth]{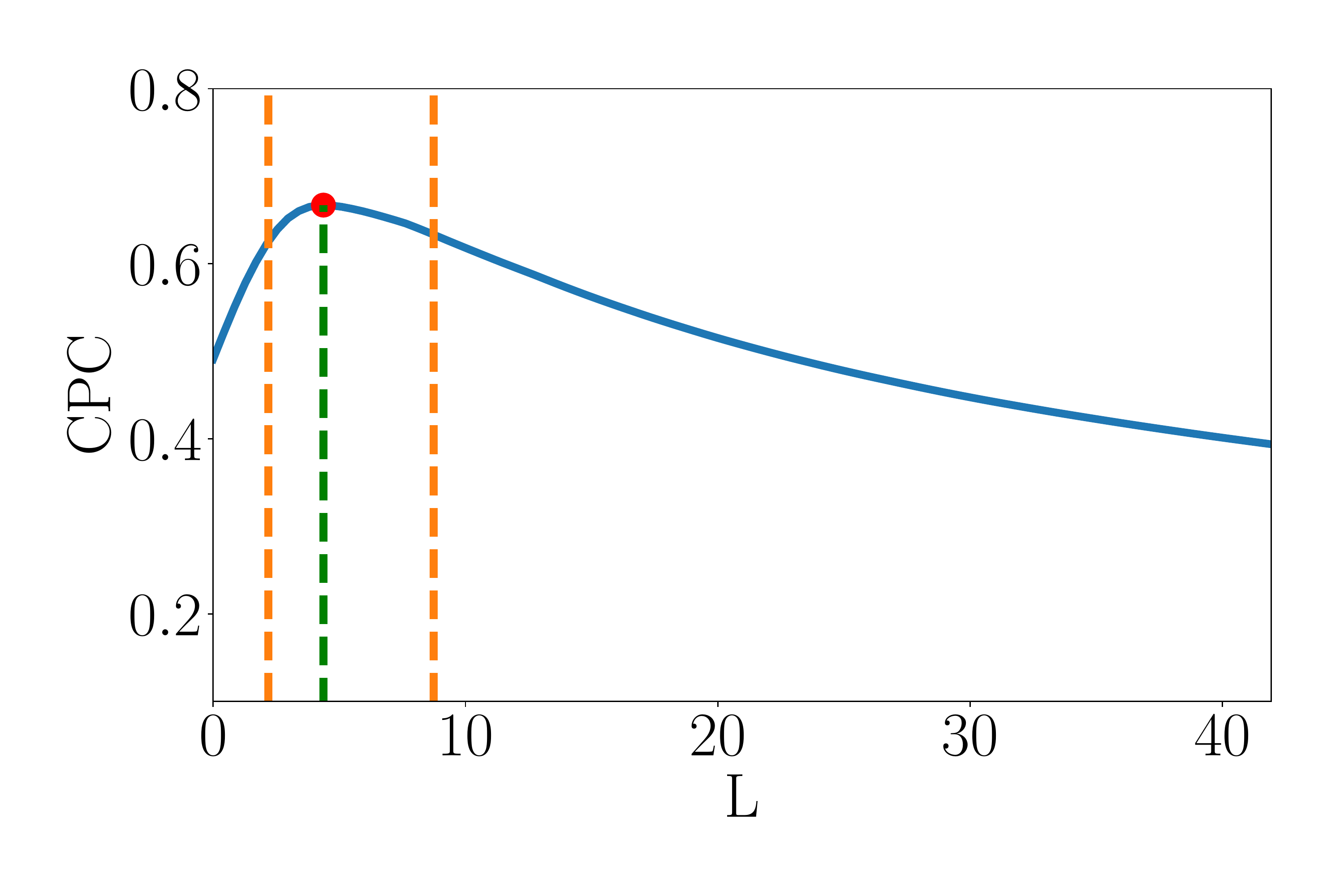}
                  \caption{IO model}
                  \label{fig:2050_CPC_vs_parameter_IO.pdf}
            \end{subfigure}%
            \begin{subfigure}[b]{0.33\textwidth}
                  \includegraphics[trim={1.2cm 1.5cm 1.2cm 1.5cm}, clip, width=\textwidth]{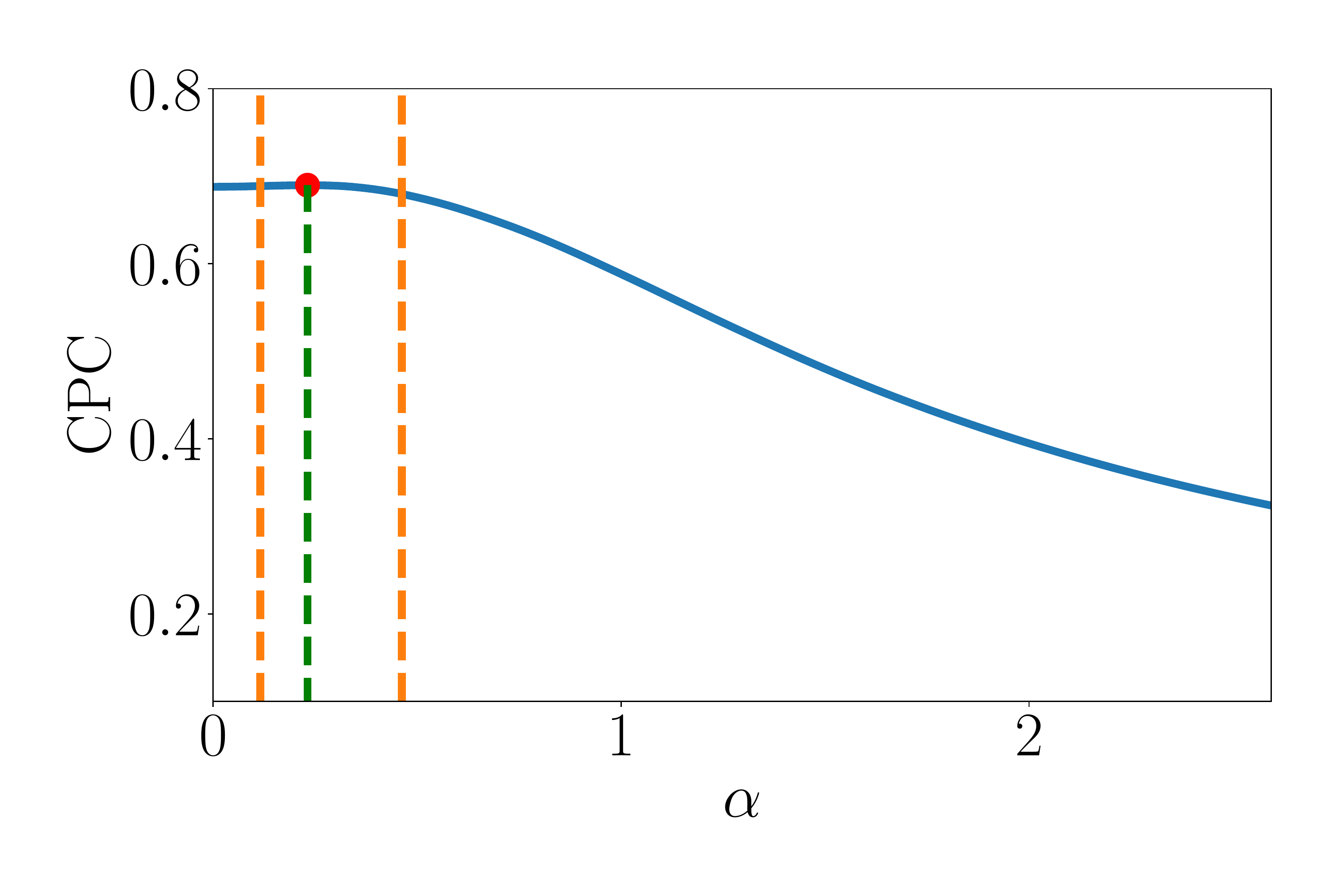}
                  \caption{Extended radiation model}
                  \label{fig:2050_CPC_vs_parameter_ext-radiation.pdf}
            \end{subfigure}
            \caption{CPC dependence on the model parameter for the gravity, IO, and extended radiation models. 
            We highlight the maximum CPC score with a red dot. The orange vertical lines are at $0.5p_{\mathrm{opt}}$ and $2p_{\mathrm{opt}}$.}
            \label{fig:CPC_vs_parameter}
        \end{figure*}

        We explore how 
        the performance of the gravity, IO, and extended radiation models depends
        on their respective model parameter values.
        For each model, let $p_{\mathrm{opt}}$ to be the optimal parameter value.
        We calculate the CPC scores for parameter values between $0$ and $10p_{\mathrm{opt}}$  (see \Cref{fig:CPC_vs_parameter}). 
        For each of the models, we observe progressively smaller CPC scores for parameter values that are progressively farther away from $p_{\mathrm{opt}}$, so model performance depends on the parameter value.
        The decrease in CPC score with distance from $p_{\mathrm{opt}}$ is steepest
        for the gravity model, second-steepest for the IO model, and shallowest for the extended radiation model.
                Interestingly, the CPC score for the extended radiation model plateaus as $\alpha \downarrow 0$ at a value close to the maximum CPC score.
        This suggests that a parameter-free 
        special case of the extended radiation model, which we obtain by setting $\alpha = 0$, may perform well.
        (We do not explore this special case
        in this article.)

    \subsection{Evaluation: Model performance on estimating trips in unseen data}
        \label{sub:estimating_trips_unseen_data}

        We conduct two series of tests to analyze the performance of the gravity, IO, and extended radiation models at estimating mobility flow for a time period for a store for which we do not know the optimal parameter value.
        As we showed in \Cref{sub:sensitivity_analysis}, the performance of each of these models depends on the value of its associated parameter.
        In each test, we estimate mobility flow using the optimal parameter value from a different time period of the same store (in our first series of tests) or from a different store for the same time period (in our second series of tests).
        In each test, we compare the achieved CPC value $\mathrm{CPC}_{\text{a}}$ with the maximum CPC value $\mathrm{CPC}_{\max}$ (which we obtain when using the optimal parameter value) with the given model. We also compute their ratio 
        \begin{equation}
            R = \frac{\mathrm{CPC}_{\text{a}}}{\mathrm{CPC}_{\max}} \in [0,1]\,.
        \end{equation}
        A value of $R$ that is close to $1$ indicates that the estimated mobility flow using a parameter value that is optimal for a different time period or for a different store fits the empirical just as successfully as using the optimal parameter value.
        In other words, these tests allow us to investigate whether the optimal parameter values of a model differ significantly between different time periods of the same store or between different stores.

        In the first series of tests, we split the $91$-day data set of each store into two parts.
        The first part is the mobility-flow data from the first $60$ days; for this subset of the data, we find the optimal parameter value for each model.
        The second part is the mobility-flow data of the remaining $31$ days; we estimate the mobility flow during this period using a mobility model with the empirical values of $O_k$ and $D_k$ (i.e., the number of OD trips that, respectively, start and end at zone $k$) from this period and the optimal parameter value from the initial $60$-day period.
        We perform one test for each store, so there are $17$ tests per model.
        The mean value of $R$ is about $0.98$ for each of the three models (see \Cref{tab:cpc_validation1}).
        Therefore, the model parameter values do not change much across different time periods of the same store.
        It thus seems that our models are likely not overfitting data, as was also noted recently for these types of mobility models by Hilton \etal \cite{hilton2019}.
        This also suggests that to estimate mobility flow of a store during some (sufficiently long) time period, we only need to know the values of $O_k$ and $D_k$ (which we can estimate from purchase data) during that time period and the optimal parameter value for a different (sufficiently long) time period of the same store.

        \begin{table}[ht]
        \caption{Mean, median, and minimum values of $R$ (i.e., the ratio of our achieved CPC value to the maximum one) for the $17$ stores in our first series of tests. In each test in this series, we use an optimal parameter value from one time period of a store to estimate mobility flow for a different time period of the same store.}
        \label{tab:cpc_validation1}
        \begin{ruledtabular}
        \begin{tabular}{l c c c}
        Model &  Mean $R$ & Median $R$& Minimum $R$\\
        \colrule
        Gravity   &  0.975  &   0.975 & 0.959\\  
        IO        &  0.978  &   0.977 & 0.963\\
        Ext. rad. &  0.975  &   0.973 & 0.960\\   
        \end{tabular}
        \end{ruledtabular}
        \end{table}

        In each test in our second series of tests, we estimate the $91$-day mobility flow of one store using a mobility model with the optimal parameter value of another store from the same time period.
        We perform one test for each 
        ordered pair of distinct stores, so there are $17 \times 16 = 272$ tests in total for each model.
        The mean value of $R$ is above $0.99$ for each of the three models (see \Cref{tab:cpc_validation2}); this suggests that the differences in model parameter values across stores are small and have minimal effect on the performance of the models.
        Therefore, we conclude that we can estimate the mobility flow of one store using the optimal parameter value from another store.
        This also suggests that if we change the 
        layout of a store, the optimal model parameter values should not change appreciably.

        \begin{table}[ht]
        \caption{Mean, median, and minimum values of $R$ (i.e., the ratio of our achieved CPC value to the maximum one) for the $272$ tests in our second series of tests. In each test in this series, we use the optimal parameter value from one store to estimate mobility flow for a different store.}
        \label{tab:cpc_validation2}
        \begin{ruledtabular}
        \begin{tabular}{l c c c}
        Model &  Mean $R$ & Median $R$& Minimum $R$\\
        \colrule
        Gravity   &  0.996  & 0.998  & 0.976 \\  
        IO        &  0.994  & 0.996  & 0.926 \\
        Ext. rad. &  0.999  & 1.000  & 0.980 \\
        \end{tabular}
        \end{ruledtabular}
        \end{table}


\section{Reducing congestion in supermarkets} \label{sec:optimization}

    We now use mobility models to estimate and reduce congestion in supermarkets.
    Our approach has three components:
     \begin{itemize}
         \item[(1)]{a congestion model, based on queuing networks, that estimates congestion from mobility flow $\bm{T}$;}
         \item[(2)]{a mobility model that estimates the change of the flow $\bm{T}$ with a new store layout; and} 
         \item[(3)]{an optimization algorithm that finds store layouts with less congestion.}
     \end{itemize}

     We describe 
     these components in detail in Sections \ref{sub:congestion_model}--\ref{sub:optimization_algorithm}.


    \subsection{Congestion model} \label{sub:congestion_model}

   In our congestion model, each node acts as a queue. We suppose that the congestion model, which resembles the one in \cite{guimera2002optimal}, is in a stationary state. (A key difference is that our model is in continuous time, whereas the model in \cite{guimera2002optimal} is discrete.) We take four inputs:
        (1) an OD matrix $\bm{T}$ with entries $T_{ij}$, which we calculate using one of the doubly-constrained mobility models in \Cref{sec:mobility_models};
        (2) a time period $\tau$ over which we measure or estimate $\bm{T}$;
        (3) a store network $\mathcal{G}$, with its associated distance matrix $\bm{\Lambda}$; and
        (4) service rates $\mu_k$ for each node $k$; as we describe later in this section, we can estimate these rates from the mean customer dwell time at $k$.

        In \Cref{sec:mobility_models}, we interpreted $\bm{T}$ as the flow from a random-walk model in which new customers arrive at a store's entrance and take trips to random destinations based on a transition matrix $\bm{P}$ whose entries are
        $P_{ij} = T_{ij} / \sum_{k} T_{ik}$ if $\sum_{k} T_{ik} > 0$ and $P_{ij} = 0$ otherwise.    
            In this section, we instead interpret $\bm{T}$ as the mean mobility flow during a time period 
        of length $\tau$ under the following model.
        New customers arrive at each node $i$ (not just at the entrance) of a network (i.e., a supermarket) according to a Poisson process with rate $\sum_k T_{ik} / \tau$. Each customer chooses a random destination $j$ with probability $P_{ij}$.
        Customers traverse the network by taking a shortest path from zone $i$ to zone $j$.
        Customers queue at each node that they visit (for both traversal and shopping) to be served,
        where each node $k$ is a single-server queue with exponential service rate $\mu_k$ \cite{kelly2011reversibility}.
        After a customer is served at the destination node $j$, we remove it from the network.
        The quantity $T_{ij}$ is then the mean number of customers who take a trip from node $i$ to node $j$ during a time period of length $\tau$.
     
        We can view the model in the formulation of the present section as a `decomposed' 
        variant of the model in \Cref{sec:mobility_models}.
        There are $n$ independent random walks, each of which starts at a different node in a network and ends after taking exactly one trip to a random destination, instead of a single random walk that always starts at the entrance node and 
        terminates at the exit node after taking one or more trips.
        In the new formulation, the mean rate at which customers finish a trip at node $k$ is the same as the mean rate at which new customers start a trip at $k$. By contrast, in the random-walk perspective of \Cref{sec:mobility_models},
        the exact number of customers who finish at $k$ is equal to the number of customers who start a trip at $k$ during any time period.
        In other words, customers are `conserved' at each node only in a stochastic sense (\ie, on average during some period of time), rather than in an absolute sense.
        
        We calculate $v_k$ for each node $k$ from $\bm{T}$ using \Cref{eq:wbc}. We need to separately consider situations with $\mu_k > v_k / \tau$ and $\mu_k < v_k / \tau$.
   
        When $\mu_k > v_k / \tau$ for all $k$, the quantity $v_k$ is the mean number of customer visits to $k$ during a time period of length $\tau$.
        The arrival rate $\lambda_k$ at each node $k$ is then $\lambda_k = v_k / \tau$.
        We call this situation a \emph{free-flow state}. In this state, the queue size at each node $k$ in stationarity is a geometric random variable with mean $\lambda_k / (\mu_k - \lambda_k)$ and is independent of the queue sizes of the other nodes \cite{kelly2011reversibility}.
        The total mean queue size $Q$ in this state is \cite{taylor2014introduction}
        \begin{equation}             \label{eq:total_mean_queue_size}
            Q = \sum_{k=1}^n \frac{\lambda_k}{\mu_k - \lambda_k}\,.
        \end{equation}

        When $\mu_k < v_k / \tau$ for some $k$, 
        node $k$ cannot serve customers sufficiently fast, and the number of customers who wait at the queue keeps increasing.
        Our system is in a
        \emph{congested state} and cannot be stationary.
     
        If we have information about the mean customer dwell time $w_k$ at each node $k$, we can infer the empirical service rate $\mu_k$ of each node $k$ using Little's Law, which states that the mean queue size is equal to the mean dwell time multiplied by the rate
        of arrivals:
        \begin{equation}
            w_k \lambda_k = q_k\,,
            \label{eq:littles_law}
        \end{equation}
        where 
        \begin{equation}   \label{eq:mean_queue_size_node}
            q_k = \frac{\lambda_k}{\mu_k - \lambda_k}
        \end{equation}
        is the mean queue size at $k$ \cite{taylor2014introduction}.           
        Combining \Cref{eq:littles_law,eq:mean_queue_size_node}, we obtain the following formula for the empirical service rate:
        \begin{equation}
            \mu_k = \frac{1}{w_k} + \lambda_k\,.
        \end{equation}

        Because we do not have empirical data for the service rate, we assume for simplicity that the service rates are homogeneous. 
        That is, $\mu_k=\mu > 0$ for all $k$, so we 
        are in a free-flow state if $\mu > \lambda_{\max}$, where $\lambda_{\max} = \max_k \lambda_k$.

        We use the maximum arrival rate $\lambda_{\max}$ and the total mean queue size $Q$ as proxies to measure congestion.
        The measure $\lambda_{\max}$, which does not depend on any parameters, 
        is the minimum service rate that ensures that the system is in a free-flow state.
        It is also closely related to the traffic capacity $\rho_c$ in the traffic-dynamics model from Arenas et al. \cite{arenas2001communication} that has been used to model traffic on transportation and communication networks \cite{chen2011traffic,arenas2001communication,guimera2002optimal,ohira1998phase,zhao2005onset,mukherjee2005phase}.
        The traffic capacity $\rho_c$ is an important performance measure of the traffic-dynamics model of Arenas et al. \cite{arenas2001communication}. It represents the maximum rate at which walkers (which, in our case, represent customers) arrive at a network from outside the system before the system reaches a congested state. In the traffic-dynamics model of \cite{arenas2001communication}, one fixes the service rate $\mu$ but varies the rate at which walkers arrive from outside the system (i.e., the \emph{external arrival rate}). By contrast, we fix the external arrival rate and vary $\mu$.

        The total mean queue size $Q$, which measures congestion in a free-flow state, is equal to the total mean number of customers in a store.
        By Little's Law \cite{little1961proof}, a store layout that minimizes $Q$ also minimizes the mean trip time.
        Unlike $\lambda_{\max}$, the total mean queue size $Q$ depends on $\mu$, so we perform separate optimizations for different values of $\mu$.

        The measures $\lambda_{\max}$ and $Q$ are correlated with each other, as 
        $Q$ is a sum that
        is dominated by the terms from nodes with large values of $\lambda_i$, so store layouts with smaller values of $Q$ often also have smaller values of $\lambda_{\max}$.


    \subsection{Mobility model} 
    \label{sub:mobility_model_congestion}          
 
        We focus on the gravity model to estimate changes in the OD matrix $\bm{T}^{\mathrm{model}}$ when changing a store's layout, because it provides the best fit to the data, both in terms of the CPC score and in the estimated number $v_k$ of visits (see \Cref{tab:cpc_rmse}).
        We assume that we can swap the locations of nodes (which corresponds to swapping the contents of their shelves), but that we cannot change the store network topology or edge distances in any other way.
        To ensure that similar items stay with one another in the same aisle, we add a further constraint (which we call the \emph{aisle constraint}) that we can only swap an aisle (which consists of a set of nodes) with another aisle with the same number of nodes.
        However, we do allow permuting of nodes within the same aisle. (In Appendix~\ref{app:SA_parameters_swap_neighbors}, we also report our results when we relax the aisle constraint. These results are of similar quality to our more constrained approach in this section.) 
        We do not consider adding or removing nodes or edges, as such changes are often costly.
        We highlight the nodes that belong to an aisle in \Cref{fig:2050_popular_nodes_original_colored_by_position_aisles.pdf} based on their color.
        Nodes of the same color are in the same aisle, and gray nodes do not belong in any aisle.
        
        Crucially, we need a hypothesis for how $O_k$ and $D_k$ (i.e., the numbers of trips to and from a node $k$) change when we change the location of a node $k$. We assume that $O_k$ and $D_k$ depend only on the items inside a zone and not on the zone's location.
        Therefore, when we change the location of a node $k$, we assume that the node has the same values of $O_k$ and $D_k$ in the new location.
        In other words, we are assuming that the number of shopping visits at node $k$ (\ie, the number of times that customers visit $k$ to purchase items) is independent of its location. This is a key assumption of our model. For nodes with many essential items, such as bread and milk, this assumption seems justifiable, as customers buy such items regardless of their location in a store. However, we anticipate that this assumption breaks down for nodes 
        in which most items are either less essential or purchased with less (or no) planning.


    \subsection{Optimization algorithm} 
    \label{sub:optimization_algorithm}
                
        We use a simulated-annealing (SA) algorithm \cite{kirkpatrick1983optimization} to find permuted layouts of a store with smaller values of one of the two objective functions ($\lambda_{\max}$ or $Q$).
        Our SA algorithm swaps 
        two aisles, which we choose uniformly at random from all pairs of aisles with the same number of nodes and whose centroids are less than $25\,\si{\metre}$ apart. After swapping the 
        two aisles, we permute the nodes within each aisle, where we choose the permutation uniformly at random from all possible permutations.
        We list the parameters of the SA algorithm in Appendix~\ref{app:SA_alg_parameters}.


    \subsection{Optimization results}
        
        We optimize a store's layout (specifically, the layout of Store A) with the SA algorithm for three examples, in which we minimize
        \begin{enumerate}
            \item[(1)] $\lambda_{\max}$;   
            \item[(2)] $Q$ with $\mu=7500$; and
            \item[(3)] $Q$ with $\mu=15000$.
        \end{enumerate}
        For $\mu = 7500$, each node serves incoming customers at a rate of $7500$ customers per day, which amounts to $12.5$ customers per minute in a store that is open for $10$ hours.
        The maximum arrival rate $\lambda_{\max}$ in the original store layout is $6575$, so a service rate of $\mu=7500$ is an example in which the most popular nodes have a longer mean dwell time.
         For example, in a store that is open for $10$ hours, the most popular node has a mean dwell time of about $38$ seconds (which we calculate using \Cref{eq:littles_law}).        
        Our example with $\mu=15000$, a value much larger than the original $\lambda_{\max}$, corresponds to a scenario in which customers typically have short mean dwell times in all nodes.
        (In this case, the most popular node has a mean dwell time of about $4.2$ seconds.)        
        We perform each optimization $20$ times and report our results in \Cref{tab:SA_results}. 
        For all three examples, the SA algorithm produces store layouts with objective-function values that are significantly smaller than their values in the original store layout. The relative reduction in $Q$ is smaller for $\mu=15000$ than it is for $\mu=7500$.
        This is not surprising, as with $\mu=7500$, a larger fraction of the total mean queue size $Q$ comes from nodes with the highest arrival rates in the original network. For example, the sum of the mean queue sizes of the three nodes with the highest arrival rates constitute
        39\% of $Q$ for $\mu=7500$. 
        By contrast, these three nodes contribute only 15\% of the value of $Q$ for $\mu=15000$.
        Therefore, store layouts that lower the arrival rates of the most congested nodes (while the arrival rates of the other nodes do not increase much) tend to have lower values of $Q$ for $\mu=7500$.
        For $\mu=15000$, because the most congested nodes contribute less to $Q$ than for they do for $\mu=7500$, 
        achieving a major relative reduction in $Q$ requires
        reducing the arrival rates of a larger number of nodes. 
        This is potentially a very difficult task.
        Therefore, our observation of a lower reduction in $Q$ for $\mu=15000$ than for $\mu=7500$ is consistent with expectations.

        \begin{table}[ht]
        \caption{Minimum and mean values of objective functions of the final
        store layouts from $20$ runs of the SA algorithm for optimizing Store A. 
        For each objective function, we show the original value of the objective function, its minimum final value across $20$ runs of the optimization algorithm, and its mean final value across the $20$ runs.}
        \label{tab:SA_results}
        \begin{ruledtabular}
        \begin{tabular}{l @{\hspace{0.2cm}} l l l}
        Objective function & Original & Minimum value & Mean value\\
        \colrule
        $\lambda_{\max}$            & 6575 & 5009  ($-23.8\%$) & 5042 ($-23.3\%$) \\
        $Q$ (with $\mu = 7500$)     & 38.10 & 29.10  ($-23.6\%$) & 29.28 ($-23.1\%$) \\
        $Q$ (with $\mu = 15000$)    & 12.78 & 11.86  ($-7.2\%$) & 11.93 ($-6.7\%$) \\
        \end{tabular}
        \end{ruledtabular}
        \end{table}
        %

        \begin{figure*}
            \begin{subfigure}[b]{0.99\columnwidth}
                  \includegraphics[trim={9cm 6cm 8cm 8cm},clip,width=\textwidth]{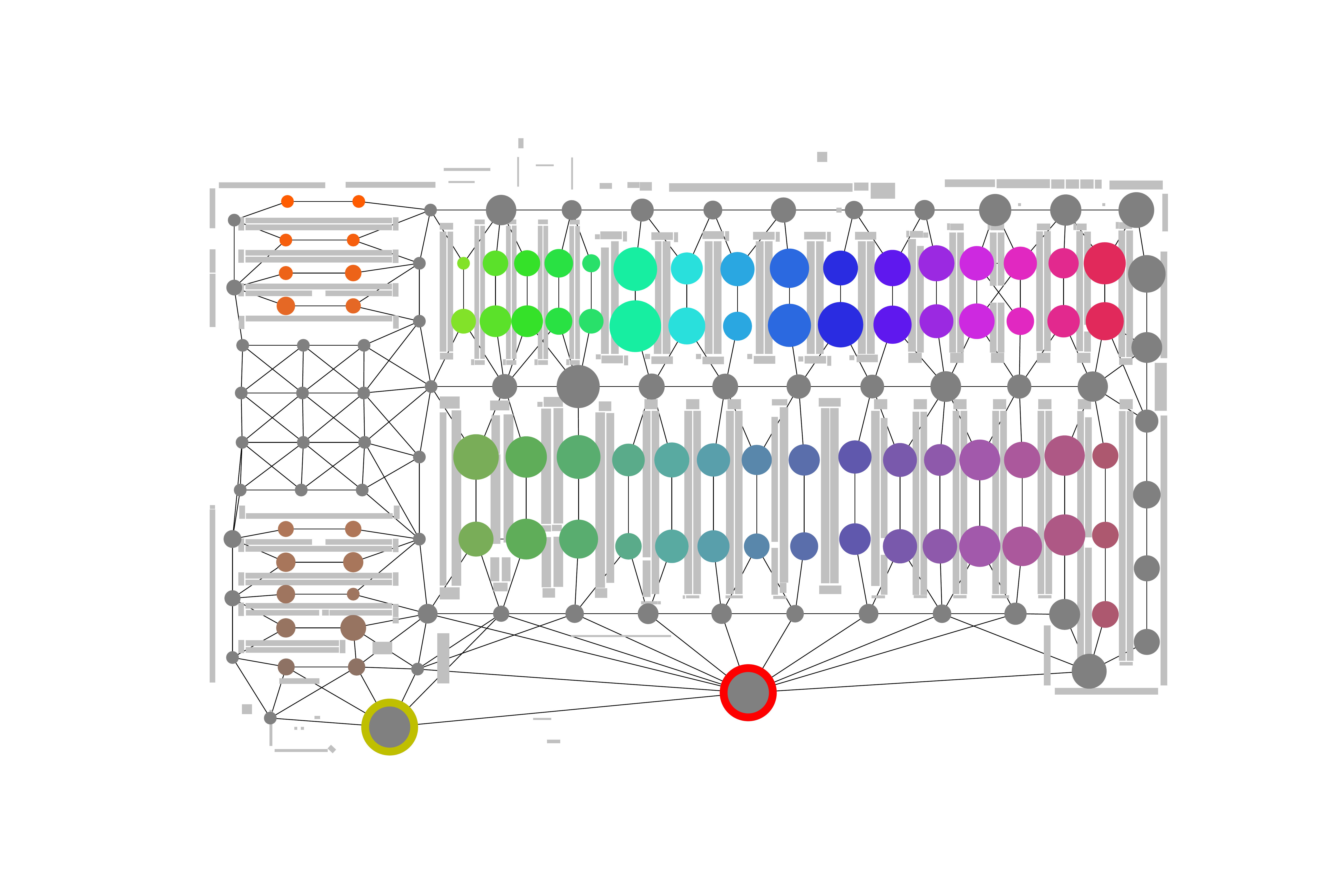}
                  \caption{Original store layout}
                  \label{fig:2050_popular_nodes_original_colored_by_position_aisles.pdf}
            \end{subfigure}%
            \begin{subfigure}[b]{0.99\columnwidth}
                  \includegraphics[trim={9cm 6cm 8cm 8cm},clip,width=\textwidth]{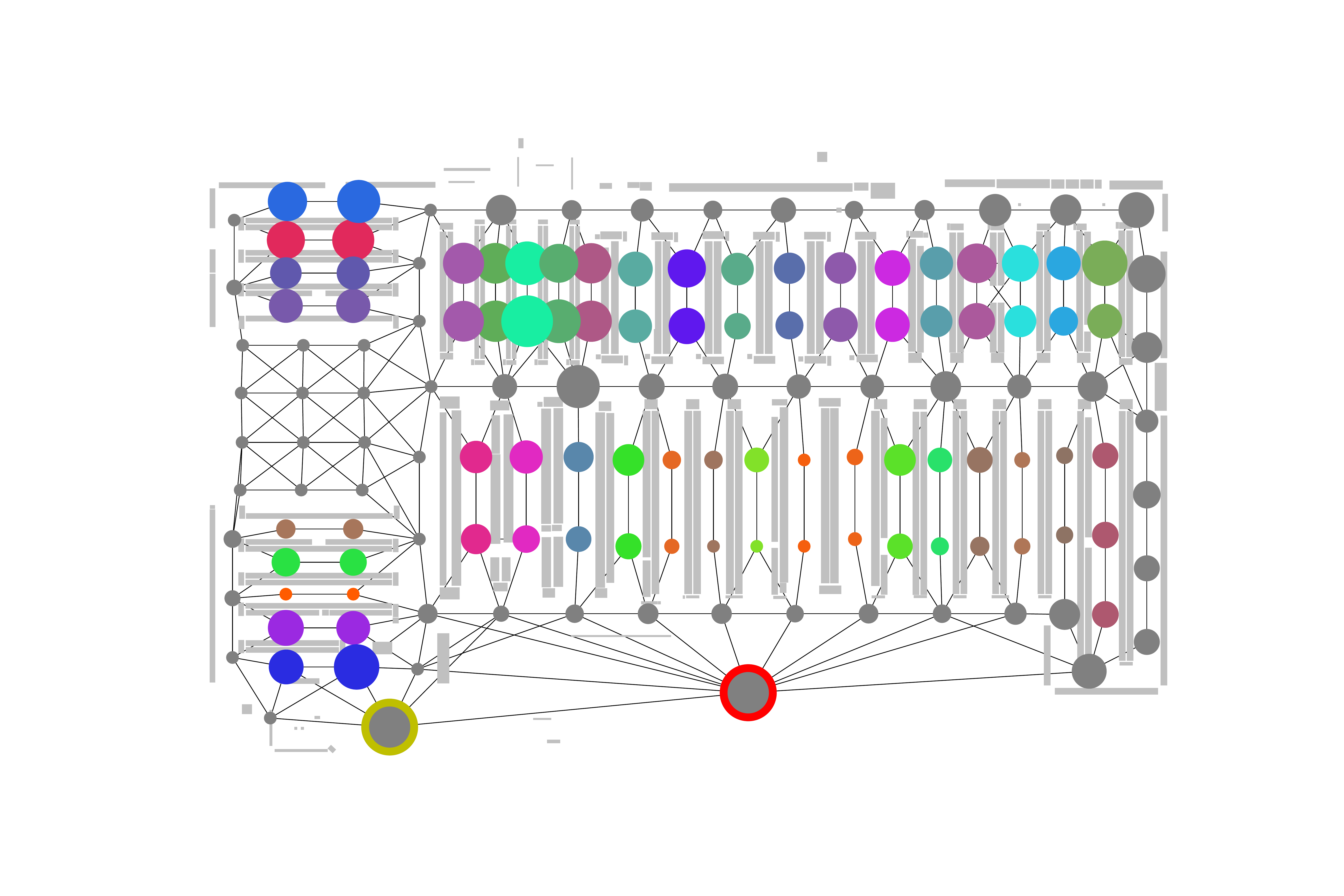}
                  \caption{Optimized store layout when minimizing $\lambda_{\max}$}
                  \label{fig:2050_popular_nodes_aisles_gravity-pl_max_arrival_rate__5009_colored_by_position_aisles.pdf}
            \end{subfigure}
            \par\bigskip 
            \begin{subfigure}[b]{0.99\columnwidth}
                  \includegraphics[trim={9cm 6cm 8cm 8cm},clip,width=\textwidth]{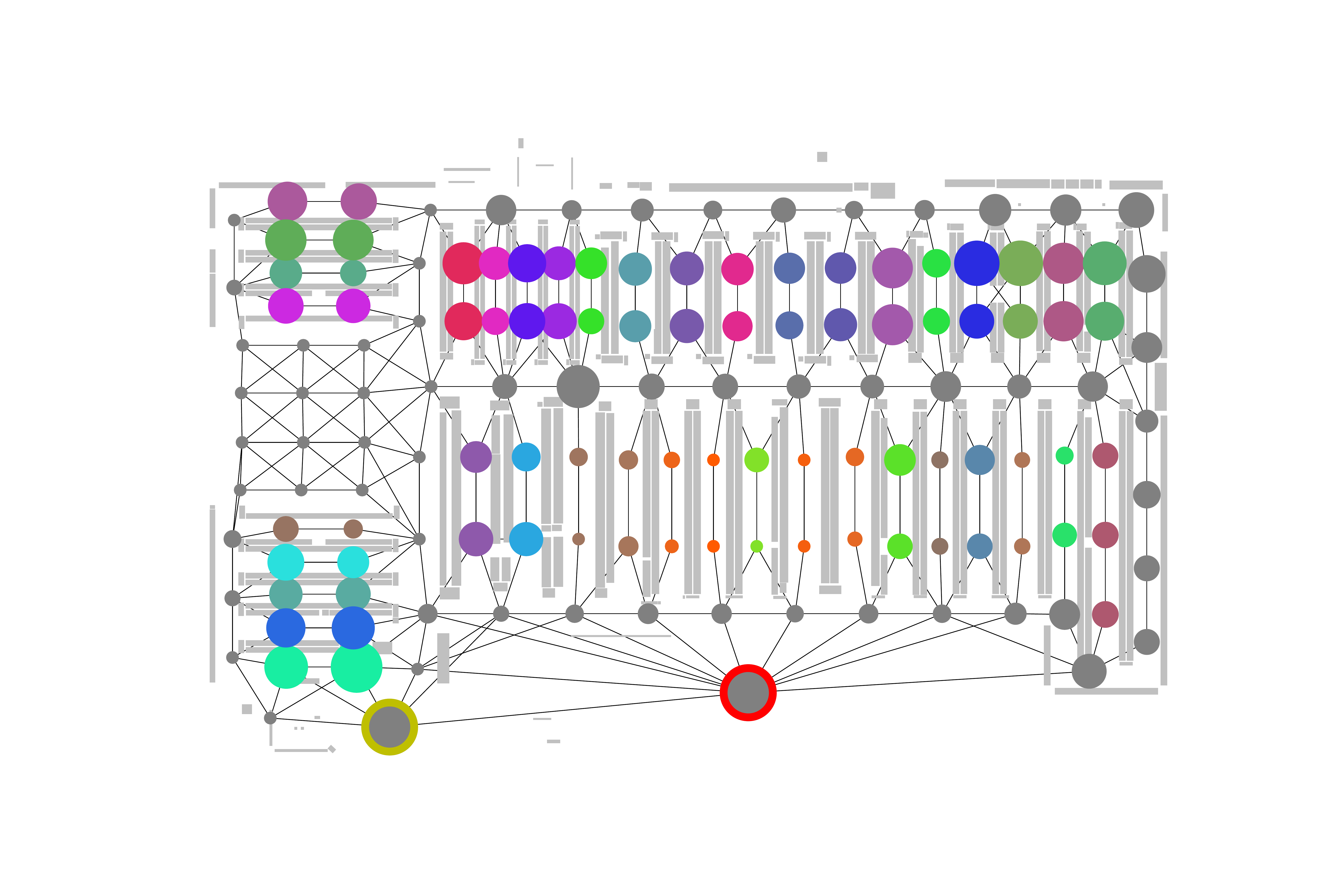}
                  \caption{Optimized store layout\\ when minimizing $Q$ (with $\mu=7500$)}
                  \label{fig:2050_popular_nodes_aisles_gravity-pl_total_mean_queue_size_7500_29.099_colored_by_position_aisles.pdf}
            \end{subfigure}
            \begin{subfigure}[b]{0.99\columnwidth}
                  \includegraphics[trim={9cm 6cm 8cm 8cm},clip,width=\textwidth]{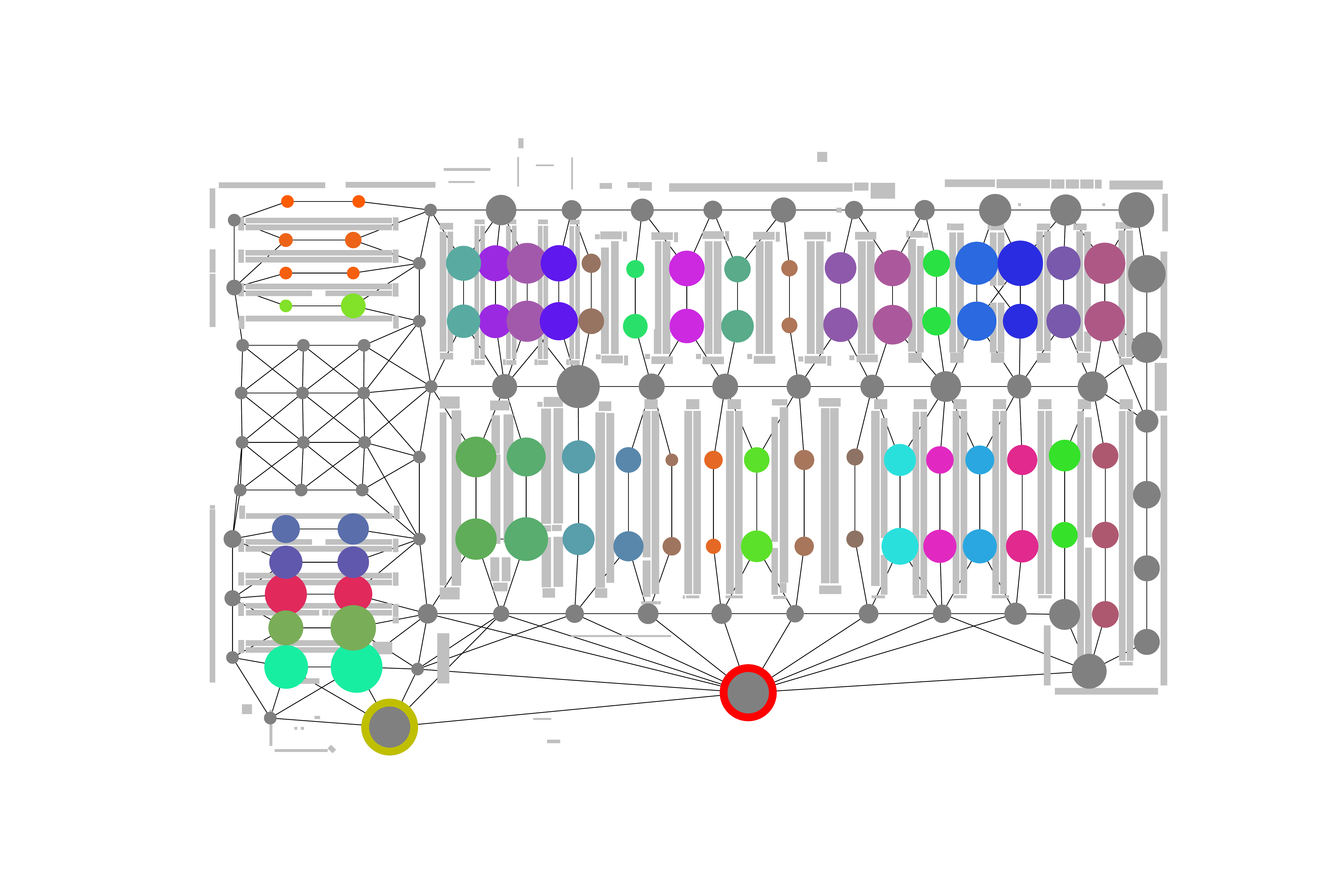}
                  \caption{Optimized store layout\\ when minimizing $Q$ (with $\mu=15000$)}
                  \label{fig:2050_popular_nodes_aisles_gravity-pl_total_mean_queue_size_15000_11.860_colored_by_position_aisles.pdf}
            \end{subfigure}
            \caption{Location of popular nodes before and after optimization. Nodes of the same color belong to the same aisle, and gray nodes do not belong to any aisles. The size of each node $k$ is proportional to $O_k + D_k$ (i.e., to the sum of the numbers of trips that start and end at zone $k$).
            We circle the entrance and till nodes in yellow and red, respectively.}
            \label{fig:popular_nodes_optimization}
        \end{figure*}

        We measure the popularity of node $k$ by $O_k + D_k$ (i.e., by the sum of the numbers of trips that start and end at $k$). For each node except the entrance and till nodes, $O_k + D_k$ is equal to twice the number of shopping visits to that node.
        In the networks with the smallest $Q$ with $\mu=7500$ and $\lambda_{\max}$, our optimization tends to move popular nodes from the center of Store A towards the left and top of the store (see \Cref{fig:2050_popular_nodes_aisles_gravity-pl_max_arrival_rate__5009_colored_by_position_aisles.pdf}). 
        By contrast, when $\mu=15000$, many popular nodes remain in the center of Store A (see \Cref{fig:2050_popular_nodes_aisles_gravity-pl_total_mean_queue_size_7500_29.099_colored_by_position_aisles.pdf,fig:2050_popular_nodes_aisles_gravity-pl_total_mean_queue_size_15000_11.860_colored_by_position_aisles.pdf}). However, our optimization moves some of them to the store's bottom-left area, which previously was not a popular area.


\section{Conclusions and discussion}
 \label{sec:discussion_and_conclusion}

    We employed several population-level mobility models to investigate customer mobility flow between zones in supermarkets, whose spatial scales are much smaller than in previous uses of these models. We estimated origin--destination (OD) matrices, which describe empirical mobility flow, for $17$ supermarkets from anonymized and ordered customer-basket data (where a customer's ``OD trip'' is either a trip between consecutive purchases, a trip from the entrance to the first purchase, or a trip from the last purchase to the tills). 
    We fit the mobility models to empirical distributions of customer {OD} trips and examined the adjustment of store aisles to reduce congestion in supermarkets. 

    Among the models that we studied, the gravity model gave the best fit to the empirical mobility flow (it successfully estimated about 69\% of the flow on average), and the extended radiation and intervening-opportunities (IO) models were almost as successful.
    This illustrates that one can successfully use population-level mobility models for applications on spatial scales of tens to hundreds of meters.
 
    In our investigation, we estimated the number $v_k$ of visits to each node $k$ from mobility flow by assuming that each customer traverses a shortest path, and we found that our estimations from the origin--destination (OD) matrices from the gravity, IO, and extended radiation models agree well with the total number of visits that we estimated from empirical OD matrices.
    (We used the shortest-path assumption only to estimate the number of visits; our estimates of mobility flow do not rely on this assumption.)
    Additionally, the gravity, IO, and extended radiation models yield trips with similar distance distributions to the empirical distribution.
    However, consistent with other studies on small spatial scales (which generally have been in intra-urban settings) \cite{lenormand2012universal,masucci2013gravity,liang2013unraveling}, the basic radiation model was not successful at reproducing features of the data.

    The gravity, IO, and extended radiation models each have one parameter, and their performance depends on the value of their parameter.
    In our investigation, we found that it is sufficient to use the optimal model parameters that we calibrated on a single store to give good estimates of the mobility flows of all other stores.
    The only additional information that we needed for the other stores is the number of trips from and to each node; one can estimate these quantities from the purchase data of these stores. 
    For a given store, we were also successful at using the models to estimate the mobility flow of a time period using a parameter value from fitting to data for another time period of the same store.
    Given our success at translating optimal parameter values across both stores and time periods, our approach provides a potentially valuable testbed for experimentation by supermarket companies using sales data from existing stores before trying out new store layouts.

    Finally, we showed how to use the gravity model in conjunction with a congestion model --- with tests using congestion measured based either on the maximum number $\lambda_{\max}$ of visits or on the total mean queue size $Q$ --- and an optimization algorithm to reduce congestion in supermarkets.
    We considered a congestion model in which each node acts as a queue with service rate $\mu$, assumed that customers traverse a shortest path between two nodes, and explored the space of store layouts in which one can permute aisles (but one cannot permute individual store zones, except within the same aisle). We then used the gravity model to estimate how mobility-flow changes from permutations of a store layout. 
    In the layouts that we obtained by minimizing $\lambda_{\max}$ or minimizing $Q$ with low service rate $\mu$, popular nodes (as measured by the number of trips from and to a node) move from the center of a store to the left and upper perimeters. 
    By contrast, in the layouts that we obtained by minimizing $Q$ with a high service rate $\mu$, some popular nodes move to a previously unpopular corner of a store.

    There are several ways to build on our work. 
    Possibilities include further development of mobility models and congestion models, analyzing seasonal effects and customer heterogeneity, allowing service rates to be heterogeneous, exploring the effect of our choice of space discretization, and applying our approach to situations other than supermarkets. We discuss several of these items in the following paragraphs.
    
    In our investigation, we inferred empirical mobility flow from anonymized, ordered basket data of the mobility of a relatively small sample of customers (approximately 7\%) from $17$ supermarkets. Naturally, this sample also has certain biases, as our data consists primarily of baskets from regular customers. It is likely that these customers possess better knowledge than other customers of the stores in which they shop (given that they do so regularly), so their mobility patterns may not be representative of all customers of a given store.

    We have also neglected temporal information and seasonal effects in our data by aggregating the mobility flow over $\tau=91$ days. 
    However, we expect mobility flow to be different at different times of the day (and on weekdays versus weekends) and at different times of a year (e.g., during certain holidays). We also expect different zones of a store to be the most congested ones at different times.
    Given sufficient data, one can apply mobility models to data that is segmented by the time of day or by the day of a year and then compare the parameter values from independent fitting to data in different time periods.

    To study congestion, we used a simple routing and congestion model (using queues in each zone of a store). 
    We assumed shortest-path routing, but some researchers have noted that customers deviate from shortest paths between purchases \cite{hui2009research,hui2009path}.
    It is important to improve understanding of the routes that customers take between purchases.
    One possible approach is to use anonymized customer-trajectory data to develop and calibrate a stochastic routing model (e.g., using a variant of a random walk, perhaps with probabilities affected by heterogeneous fitness values for different zones of a store).
    One can incorporate such a routing model in a straightforward way into our framework to better estimate the number $v_k$ of visits to each zone $k$.
    When we estimated the mobility flow of different store layouts, we assumed that the number of customer shopping visits at node $k$ is fixed and does not depend on the location of $k$.
    However, the location of a zone that contains items that are typically bought in an unplanned way likely affects the number of shopping visits to that zone.
    One can incorporate increasingly accurate models of 
    the number and zone distribution of shopping visits into our framework to improve estimates of the mobility flow from different store layouts.
    Additionally, more empirical research is necessary to attain a detailed mechanistic understanding of the causes of congestion in supermarkets.
    We modeled congestion as queues in a zone; if such a model is conceptually accurate,
    we can incorporate more realistic types of queues (\eg, with variable service rates or with customers who do not enter a queue if it is too long).
    One can infer service rates using a method that is analogous to what we described in \Cref{sec:optimization}, provided one possesses data on customer dwell time (or can somehow infer such times) for each zone of a store.

    Another consideration is the choice of space discretization and spatial resolution, and it is necessary to examine how such choices affect qualitative results of both mobility models and congestion models.
    (In our work, we divided each store into zones of approximately similar size, with zone lengths of about $7\,\si{\metre}$.)

    Although one can apply our approach for modeling mobility flow and congestion at any spatial scale, 
    we expect that one can implement our methodology 
    in practice in systems in which one can modify the underlying spatial structure.
    Many such applications have small spatial scales, as rewiring a small system is often a lot less costly than rewiring a large one.
    For example, when considering commuting flow, one cannot change the locations of countries or buildings.
    However, one can apply our tools for modeling mobility flow and congestion in a museum and use our optimization procedure
    to suggest better locations for the exhibits.
    Other examples include poster sessions in academic conferences and food stations in buffet restaurants.
    Applying our approach to these settings will help reveal which of our findings are specific to mobility flow in supermarkets and which ones apply more generally to human mobility on small spatial scales. 


\begin{acknowledgements}

FY acknowledges support from EPSRC Centre For Doctoral Training in Industrially Focused Mathematical Modelling (EP/L015803/1) in collaboration and Tesco PLC. MBD acknowledges support from Oxford--Emirates Data Science Lab. We also thank David Allwright, Robert Armstrong, Chico Camargo, Albert D\'{\i}az-Guilera, Scott Hale, Renaud Lambiotte, Neave O'Clery, Trevor Sidery, and an anonymous referee for helpful comments. 

\end{acknowledgements}


\appendix

\section*{Appendix}


\section{Defining zones in a store}\label{app:store_zoning}

    In this appendix, we discuss how we define the zones in a store.
        
    First, we manually identify the aisles in a store.
    Each aisle is a long rectangle between two lines of shelves.
    The edges of each aisle are either parallel to the horizontal axis or parallel to the vertical axis on a floor plan.
    (In other words, there are no tilted aisles.)
    
    We then subdivide each aisle into $Z$ rectangular zones of equal dimensions, where we choose the number $Z$ of zones such that the 
    length of the zones in the aisle is as close as possible to $7\,\si{\metre}$ (our targeted mean zone length).
    (We define the ``length'' of a zone as the longer side of the rectangle that encloses the zone.)
    The number $Z$ of zones in each aisle varies between different aisles; it is typically either $2$ or $3$.
    The width of each aisles is smaller than $7\,\si{\metre}$, so if an aisle is horizontal (\ie, the longer edges of the aisle are parallel to the horizontal axis), then all zones in the aisle also horizontal. 
    Similarly, if an aisle is vertical (\ie, the longer edges of the aisle are parallel to the vertical axis), all zones are also vertical.
 
    We then manually place the entrance zone and the till zone. 
    We fill the remaining floor area of a store with long, large rectangles whose widths are similar to the width of store aisles. We divide each of these large rectangles into zones of equal size (where zones for
    different large rectangles may have different sizes).
    See \cite{ying2019PhDThesis} a lengthier discussion of how we define zones.


\section{Iterative proportional-fitting procedure for determining $A_i$ and $B_j$}
\label{app:iterative_proportional_fitting}

    In the doubly-constrained models, we are given $O_i$, $D_j$, and $f_{ij}$ for all nodes $i$ and $j$. We seek to determine $A_i$ and $B_j$ (the balancing factors) that satisfy
    \begin{align}
        O_i &= \sum_j T_{ij}^{\mathrm{model}} = \sum_j A_i O_i B_j D_j f_{ij}\,, \label{app_eq:O_i}\\
        D_j &= \sum_i T_{ij}^{\mathrm{model}} = \sum_j A_i O_i B_j D_j f_{ij}\,. \label{app_eq:D_j}
    \end{align}
    Rearranging \Cref{app_eq:O_i,app_eq:D_j} yields
    \begin{align}
        A_i &= \left(\sum_j B_j D_j f_{ij}\right)^{-1} \label{app_eq:A_i}\,,\\
        B_j &= \left(\sum_i A_i O_i f_{ij}\right)^{-1} \label{app_eq:B_j}\,.
    \end{align}
    In our iterative proportional-fitting procedure, we initialize $A_i = 1$ for all $i$.
    We then calculate $B_j$ using \Cref{app_eq:B_j} from $A_i$, followed by an update of $A_i$ using \Cref{app_eq:A_i}.
    We repeat this procedure until the values on the right-hand side of \Cref{app_eq:O_i,app_eq:D_j} are close (specifically, within 1\%) of the values on the left-hand side. In practice, this procedures converges within $1000$ iterations for all of the employed models.


\section{Performance of the doubly-constrained gravity model with an exponential deterrence function} \label{app:exp_deterrence}

    In the doubly-constrained gravity model with an exponential deterrence function, the OD matrix $\bm{T}^{\mathrm{model}}$ is given by \Cref{eq:base_model} with 
    \begin{equation}
          f_{ij} = f_{\mathrm{g}}(O_i, D_j, d_{ij}) = O_i D_j e^{-\gamma d_{ij}/l}\,.
      \label{eq:gravity_exp}
    \end{equation}

    We find that both the CPC score and NRMSE$_v$ are slightly worse on average than for the gravity model with a power-law deterrence function (see \Cref{tab:cpc_rmse_gravity_model-specific}).

    \begin{table}[ht]
        \caption{Mean CPC scores and $\mathrm{NRMSE}_v$ when fitting the doubly-constrained gravity model with an exponential deterrence function versus the doubly-constrained gravity model with a power-law deterrence function.}
        \label{tab:cpc_rmse_gravity_model-specific}
        \begin{ruledtabular}
            \begin{tabular}{l c c c c}
            Model & Mean CPC & Mean $\mathrm{NRMSE}_v$  \\
            \colrule
            Gravity (exponential)  & 0.677 & 0.049 \\  
            Gravity (power law)  & 0.686 & 0.045 \\  
            \end{tabular}
        \end{ruledtabular}
    \end{table}


\section{Estimating $\{O_k\}_{k=1}^n$ and $D_1^{\mathrm{data}}$ from purchase data} 
\label{app:shopping_visits}

    In our models, we used $\{O_k\}_{k=1}^n$ and $D_1^{\mathrm{data}}$ to calculate mobility flow. In the main text, we assumed that we know these values. 
    In this section, we show how to estimate $\{O_k\}_{k=1}^n$ and $D_1^{\mathrm{data}}$ from customer-level purchase data.

    For each customer $c$, our data includes a list of items that were purchased during a shopping journey.
    Using item-location data, we can identify the possible zones in which a customer could have picked up each item.
    We assume that all items 
    in the same zone were picked up in one visit by a customer, so customers do not visit a zone more than once.
    The main challenge is how to account for purchased multi-located items.
    We calculate the number $O_k^{\mathrm{data}, c}$ of trips that start from nodes $k=1,\dots,n$ for each customer $c$ as follows. 
    If a customer buys an item that is located only in zone $k$, we set $O_k^{\mathrm{data}, c} = 1$. 
    Otherwise, we check whether a customer buys an item that is located both in zone $k$ and in other zones. (In other words, there are multiple possible zone locations for that item.) If this is the case, let $M$ be the number of such purchased items, and let $N_1, \dots, N_M$ be the number of possible zone locations for each of the $M$ items. We 
    set $O_k^{\mathrm{data},c} = \min \left\{\sum_{l=1}^M 1/N_{l}, 1 \right\}$.
    Therefore, each item that is located in zone $k$ and in $N-1$ other zones counts as a $1/N$ of a visit, with $O_k^{\mathrm{data},c}$ capped at $1$.
    If a customer has not purchased any items that are located in zone $k$, then $O_k^{\mathrm{data},c} = 0$.
    By construction, the value of $O_k^{\mathrm{data},c}$ is at most $1$, and it can take a fractional value when a customer buys items that are located in $k$ as well as other zones.

    \begin{figure}[t]
          \includegraphics[width=0.65\columnwidth]{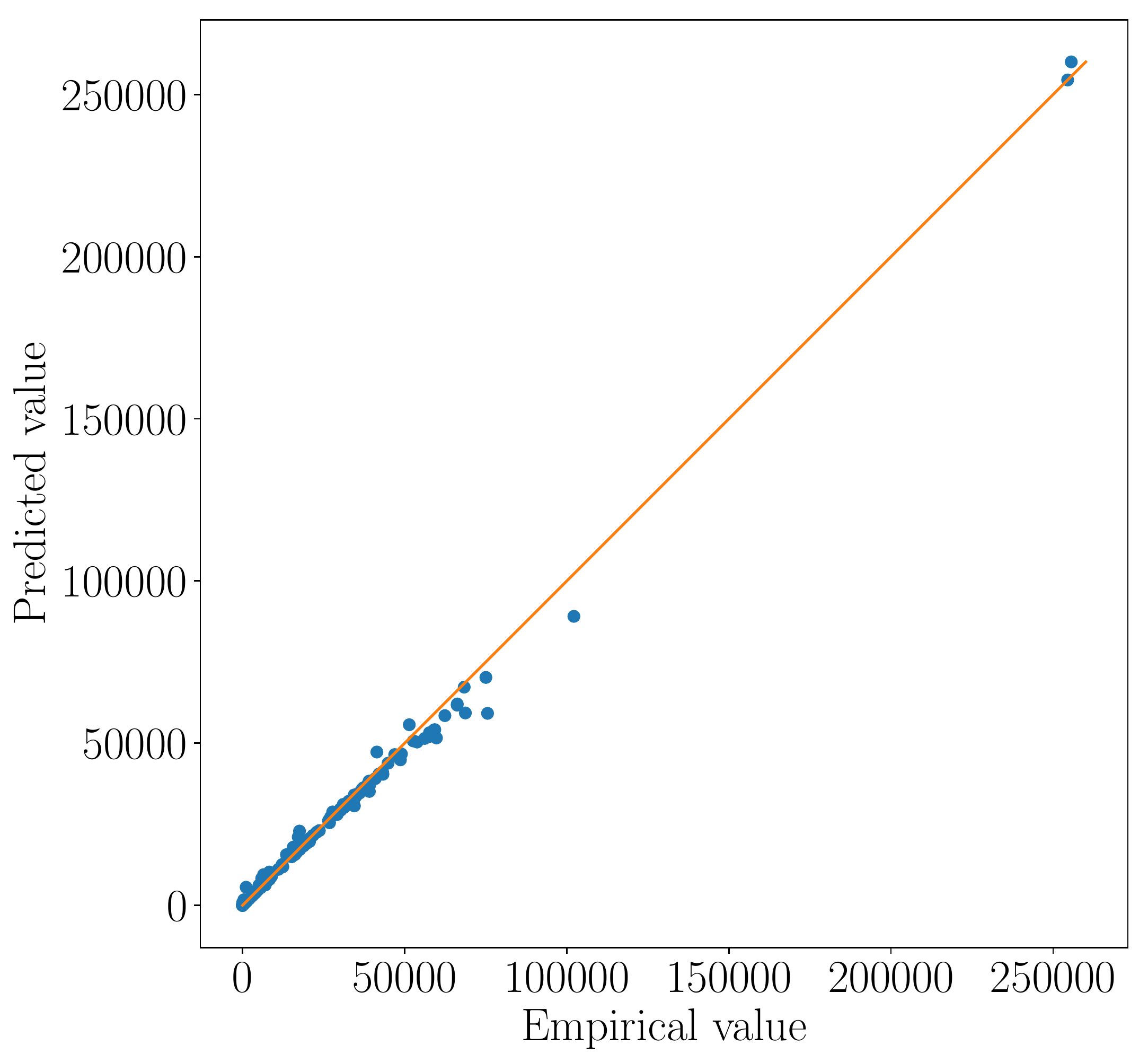}
        \caption{Comparison of our estimates and empirical values of $O_k$, $D_1^{\mathrm{data}}$, and $D_n^{\mathrm{data}}$ for Store A.
        }
        \label{fig:2050_shopping_visits_prediction.pdf}
    \end{figure}

    We calculate $O_k$ (with $k=1,\dots,n$) with the formula
    \begin{equation}
        O_k = \sum_c O_k^{\mathrm{data},c} + \delta_{k1} C\,,
    \end{equation}
    where $C$ is the number of journeys in the data set and $\delta_{k1}$ is the Kronecker delta.
    The term $\delta_{k1} C$ accounts for the first trips of each shopping journey, as these start at the entrance of a store. 
    We calculate $D_1^{\mathrm{data}}$ with the formula
    \begin{align}
        D_1^{\mathrm{data}} &= \sum_c O_1^{\mathrm{data},c}\,.
    \end{align}
    We rescale $\{O_k\}_{k=1}^n$, and $D_1^{\mathrm{data}}$ by dividing by $\rho$ (i.e., by the number of baskets in the data set during the time period divided by the total number of baskets during the time period).

    In \Cref{fig:2050_shopping_visits_prediction.pdf}, we compare the empirical values of $\{O_k\}_{k=1}^n$ and $D_1^{\mathrm{data}}$ with ones that we estimate for Store A. In our calculations, the estimated values are close to the empirical ones, so we conclude that we can estimate $O_k$ and $D_k$ using only purchase data.


\section{Derivation of the attraction factor $f_{ij}$ in Schneider's IO model}
    \label{app:derivation_schneider_IO_model}
    
    In Schneider's IO model, each customer who leaves node $i$ considers each opportunity in nondecreasing order of distance from $i$ (\ie, from the nearest node to the farthest one).
    For simplicity, we assume that there are no equidistant nodes. A customer accepts the current opportunity with probability $L/N$, where 
    $L \in (0, N]$ is a dimensionless fitting parameter and $N=\sum_i O_i^{\mathrm{data}}$ is the total number of trips~\footnote{In the original formulation, $L/N$ is replaced by $L$ in \Cref{eq:io_model_base}, and $L$ has dimensions of [(number of trips)$^{-1}$]}.
    Upon accepting an opportunity at $j$, a customer takes a trip from node $i$ to node $j$ and thus does not consider any other
    opportunities.
    A customer who has not accepted any opportunity restarts the process and considers all opportunities again in nondecreasing distance from $i$.
    This process continues until the customer accepts an opportunity. 
    
    We calculate the probability of a customer accepting an opportunity at node $j$ as follows.
    Let $M$ be the number of rejected opportunities in the final iteration when an opportunity is accepted.
    The random variable $M$ has a truncated geometric distribution.
    A customer who accepts an opportunity at node $j$ must reject all $S_{ij}$ opportunities at closer nodes and accept one of the $D_j$ opportunities at $j$. In other words, a customer in zone $i$ takes a trip to zone~$j$ if $S_{ij} + D_j > M \geq S_{ij}$.
    The probability of rejecting at least $k$ opportunities is
    \begin{equation}
       \mathbb{P}(M \geq k) \propto \left(1 - \frac{L}{N}\right) ^ k \approx \exp\left(-\frac{kL}{N}\right)\,, \quad k < N \,,   
    \end{equation}
    where the approximation becomes exact as $N$ and $k$ tend to infinity with $k/N$
    constant.
    Therefore, the probability that a customer accepts an opportunity at node $j$ is
   \begin{widetext}
    \begin{equation} \label{e2}
    \begin{aligned}
        \mathbb{P}(S_{ij} + D_j > M \geq S_{ij}) 
        = \mathbb{P}(M \geq S_{ij}) - \mathbb{P}(M \geq S_{ij} + D_j) 
         \approx e^{-\frac{L}{N} S_{ij}} - e^{-\frac{L}{N} (S_{ij} + D_j)} = f_{ij}\,.
    \end{aligned}
    \end{equation}
    \end{widetext}
    In Equation \eqref{e2}, the quantity $f_{ij}$ equals the number of customers who make a trip from $i$ to $j$ divided by the number of customers who leave $i$. 
    However, as we use a doubly-constrained model, $f_{ij}$ gives only the attraction value of zone $j$ to a customer in zone $i$. 
    {In such a model, the quantity $T^{\text{model}}_{ij} / O_i = A_i B_j D_j f_{ij}$ equals the actual number of customers who make a trip from $i$ to $j$ divided by the total number of trips that originate at $i$.}


\section{Parameters of the simulated-annealing algorithm}
\label{app:SA_alg_parameters}

    We set the initial computational temperature to $200$ when minimizing $\lambda_{\max}$ and to $20$ when minimizing $Q$.
    We use a cooling schedule in which we reduce the temperature by 0.18\% of the current temperature at each step, so there are $5000$ steps in total.
    We use the standard acceptance probability function $\exp(-\Delta E / T)$, where $\Delta E$ is the change in the objective-function value of the current step and $T$ is the computational temperature.


\section{Results of the simulated-annealing algorithm without the aisle constraint}
\label{app:SA_parameters_swap_neighbors}

    \begin{table*}[ht]
        \caption{Minimum and mean values of objective functions of the final store layouts from $20$ runs without the aisle constraint and $20$ runs with the aisle constraint of the SA algorithm for optimizing Store A. 
        For each objective function, we show the original value of the objective function, its minimum final value across $20$ runs of the optimization algorithm, and its mean final value across the $20$ runs.}
        \label{tab:SA_results_swap_neighbors}
        \begin{ruledtabular}
        \begin{tabular}{l @{\hspace{0.2cm}} l l l l l}
                           & & \multicolumn{2}{c}{Without the aisle constraint} & \multicolumn{2}{c}{With the aisle constraint} \\
        Objective function & Original & Minimum value & Mean value & Minimum value & Mean value\\
        \colrule
        $\lambda_{\max}$            & 6575  & 5040  ($-23.3\%$) & 5150 ($-21.7\%$)  & 5009  ($-23.8\%$) & 5042 ($-23.3\%$) \\
        $Q$ (with $\mu = 7500$)     & 38.10 & 27.21  ($-28.6\%$) & 28.02 ($-26.5\%$) & 29.10  ($-23.6\%$) & 29.28 ($-23.1\%$) \\
        $Q$ (with $\mu = 15000$)    & 12.78 & 11.06  ($-13.5\%$) & 11.23 ($-12.1\%$) & 11.86  ($-7.2\%$) & 11.93 ($-6.7\%$) \\
        \end{tabular}
        \end{ruledtabular}
    \end{table*}

    We now apply the SA algorithm without the aisle constraint.
    Specifically, in one swapping step, we choose an edge uniformly at random among all edges, except for those that are incident to the entrance or till node; and (assuming the step is accepted) we swap the two nodes that are incident to chosen edge. 
    We use the same parameters that we described in Appendix~\ref{app:SA_alg_parameters}.
    The SA algorithm without the aisle constraint finds store layouts with smaller values of $Q$ on average than when we run the algorithm with the aisle constraint (see \Cref{tab:SA_results_swap_neighbors}).
    We find that the relative decrease in $Q$ is smaller for $\mu=7500$ than for $\mu=15000$.
    Interestingly, the store layouts that we obtain when minimizing $\lambda_{\max}$ have larger values of $\lambda_{\max}$ without the aisle constraint than with the constraint.
    This suggests that the SA algorithm, which is a heuristic algorithm, is not very efficient at exploring the state space of store layouts, as it is unable to find the store layouts (which have small values of $\lambda_{\max}$) that were obtained with the aisle constraints.




%


\end{document}